\documentclass{aastex6}

\begin{document}

\title{Dense molecular gas in the starburst nucleus of NGC 1808}

\author{Dragan Salak\altaffilmark{1}, Yuto Tomiyasu\altaffilmark{2}, Naomasa Nakai\altaffilmark{2,3}, Nario Kuno\altaffilmark{2,3}, Yusuke Miyamoto\altaffilmark{4}, and Hiroyuki Kaneko\altaffilmark{4}}

\altaffiltext{1}{School of Science and Technology, Kwansei Gakuin University, 2-1 Gakuen, Sanda, Hyogo 669-1337, Japan, d.salak@kwansei.ac.jp}
\altaffiltext{2}{Division of Physics, Faculty of Pure and Applied Sciences, University of Tsukuba, 1-1-1 Tennodai, Tsukuba, Ibaraki 305-8571, Japan}
\altaffiltext{3}{Center for Integrated Research in Fundamental Science and Technology,
University of Tsukuba, 1-1-1 Tennodai, Tsukuba, Ibaraki 305-8571, Japan}
\altaffiltext{4}{Nobeyama Radio Observatory, National Astronomical Observatory of Japan, 462-2 Nobeyama, Minamimaki, Minamisaku, Nagano 384-1305, Japan}

\begin{abstract}

Dense molecular gas tracers in the central 1 kpc region of the superwind galaxy NGC 1808 have been imaged by ALMA at a resolution of 1\arcsec (\(\sim50\) pc). Integrated intensities and line intensity ratios of HCN (1-0), H\(^{13}\)CN (1-0), HCO\(^{+}\) (1-0), H\(^{13}\)CO\(^{+}\) (1-0), HOC\(^{+}\) (1-0), HCO\(^{+}\) (4-3), CS (2-1), C\(_2\)H (1-0), and previously detected CO (1-0) and CO (3-2) are presented. SiO (2-1) and HNCO (4-3) are detected toward the circumnuclear disk (CND), indicating the presence of shocked dense gas. There is evidence that an enhanced intensity ratio of HCN (1-0)/HCO\(^{+}\) (1-0) reflects star formation activity, possibly in terms of shock heating and electron excitation in the CND and a star-forming ring at radius \(\sim300\) pc. A non-LTE analysis indicates that the molecular gas traced by HCN, H\(^{13}\)CN, HCO\(^{+}\), and H\(^{13}\)CO\(^{+}\) in the CND is dense (\(n_\mathrm{H_2}\sim10^5~\mathrm{cm}^{-3}\)) and warm (\(20~\mathrm{K}\lesssim T_\mathrm{k}\lesssim100~\mathrm{K}\)). The calculations yield a low average gas density of \(n_\mathrm{H_2}\sim10^2-10^3~\mathrm{cm}^{-3}\) for a temperature of \(T_\mathrm{k}\gtrsim30~\mathrm{K}\) in the nuclear outflow. Dense gas tracers HCN (1-0), HCO\(^{+}\) (1-0), CS (2-1), and C\(_2\)H (1-0) are detected for the first time in the superwind of NGC 1808, confirming the presence of a velocity gradient in the outflow direction.

\end{abstract}

\keywords{galaxies: individual (NGC 1808) --- 
galaxies: ISM --- galaxies: nuclei --- galaxies: starburst --- ISM: structure}

\section{Introduction}\label{A}

Cold molecular clouds are the sites of star formation and the densest gas phase in the interstellar medium (ISM). Sensitive observations at mm and sub-mm wavelengths, now readily feasible with telescopes such as Atacama Large Millimeter/submillimeter Array (ALMA), can yield detections of multiple molecular species from cold clouds via their rotational transitions and serve as a powerful tool to investigate the physical conditions (kinetic temperature and density) and chemical composition of the interstellar gas in our Galaxy and nearby galaxies (e.g., \citealt{Baa08,Cos11,Tak14,Vit14,Nak15,Nis16a,Nis16b,Wat17a,Wat17b}).

Exploring the physical conditions of cold gas in starburst galaxies, where star formation activity is vigorous, is particularly important, because starbursts are the main producers of feedback in the form of supernova explosions and ionizing radiation (stellar winds) from massive stars, thought to be the driving engine of superwinds (mass outflows) and a key factor in regulating star formation in galaxies (e.g., \citealt{CC85,HAM90,MQT05,VCB05,LBO17}). While the distribution of CO, as a tracer of molecular gas mass and kinematics, has been extensively studied (e.g., \citealt{Bol13,Sal13}), sensitive images of dense gas tracers with critical densities \(n_\mathrm{cr}\gtrsim10^4~\mathrm{cm}^{-3}\) (e.g., HCN, HCO\(^{+}\), CS) at \(\lesssim100\) pc resolution in starburst disks and winds are still sparse (e.g., \citealt{Sala14,Wal17}). In particular, the coexistence of dense neutral gas with hot ionized gas in extraplanar winds and its physical and chemical evolution are not fully understood. In order to probe the gas conditions in such environments, deep multi-line observations of various molecular species, including dense gas tracers, are essential.

\begin{figure}
\epsscale{0.75}
\plotone{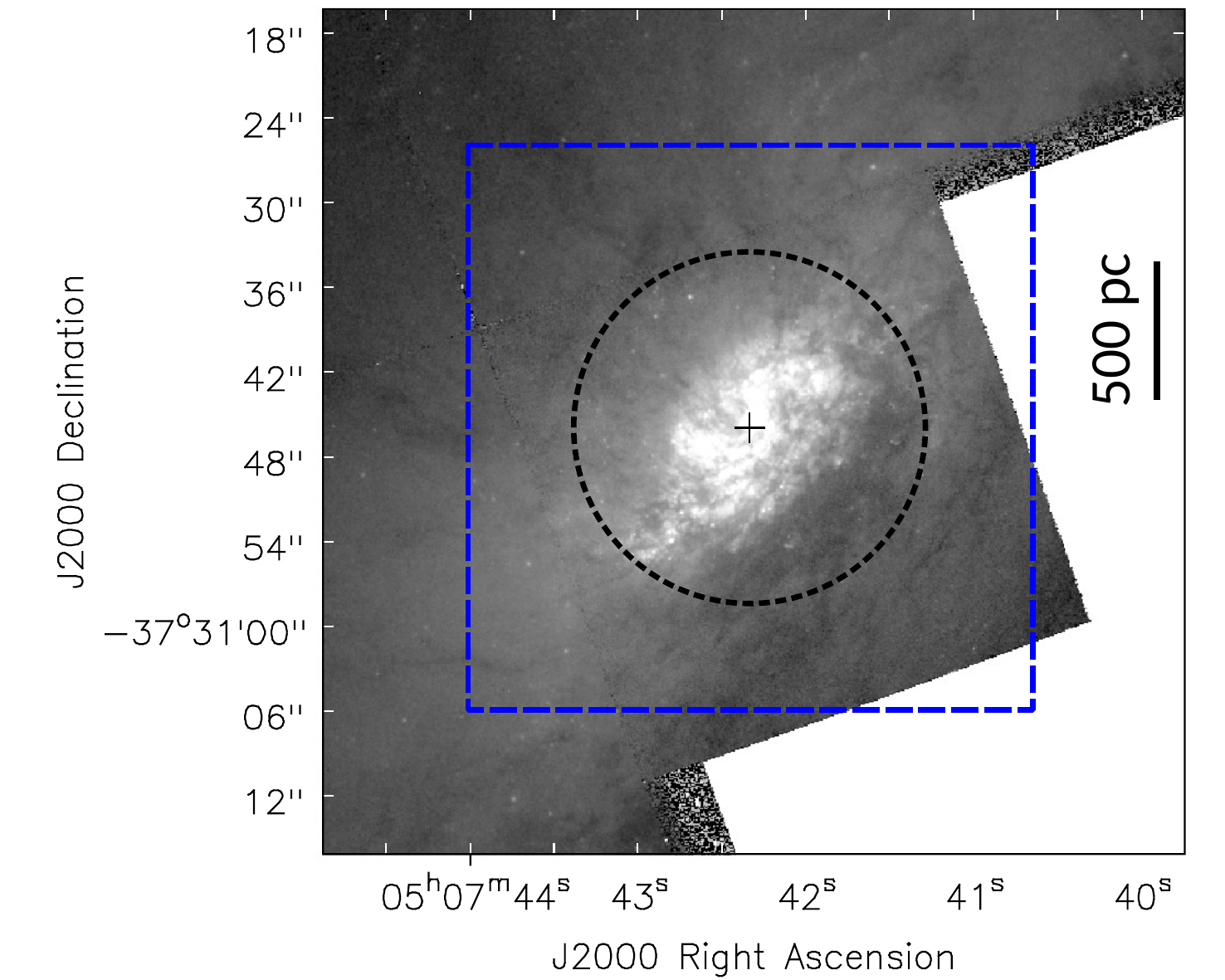}
\caption{Optical \(R\)-band image of NGC 1808 (acquired from the Hubble Legacy Archive). The blue dashed rectangle marks the size of the band 7 mapping region. The black dotted circle (diameter \(25\arcsec\)) indicates the size of the starburst nucleus. The plus sign marks the adopted galactic center derived from the 93 GHz continuum \citep{Sal17}.\label{fig:n1808}}
\end{figure}

This article is a part of a series about ALMA observations of the nearby (10.8 Mpc; \citealt{Tul88}) galaxy NGC 1808 (Figure \ref{fig:n1808}), selected as a case study to investigate the distribution of molecular gas and its physical conditions in the central starburst disk and gas outflow. The wind from the central 1 kpc region of NGC 1808 has been revealed previously in the form of polar dust lanes \citep{BB68}, detection of Na I absorption line in extraplanar space \citep{Phi93}, and peculiar kinematics (line splitting) of molecular gas traced by CO (1-0) \citep{Sal16,Sal17}. The main findings from the previous ALMA observations were: (1) molecular gas outflow from the CND and 500 pc gaseous ring, (2) a velocity gradient (possible acceleration) in the outward direction along the outflow, (3) a gradient of the line intensity ratio of CO (3-2) to CO (1-0) with galactocentric radius in the central 1 kpc, and (4) an evolutionary sequence of molecular clouds that inflow from the large-scale bar into the 500 pc ring region, trigger star formation, and get entrained into outflows \citep{Sal17}. The revealed properties make NGC 1808 a valuable nearby ``laboratory'' galaxy where we can study the nuclear gas dynamics, starburst phenomenon, and its feedback. The previous ALMA articles involved the CO and continuum data, tracing the bulk molecular gas and star formation activity, respectively. However, in order to investigate the physical conditions of the gas in the starburst disk and outflow, the next important step is to image other molecular species, especially those that trace the dense molecular gas, which is abundant in star-forming regions. In this article, we present the results of the first high-resolution ALMA observations of dense gas tracers, such as HCN (1-0), HCO\(^{+}\) (1-0), C\(_2\)H (1-0), and CS (2-1) toward the central 1 kpc region of NGC 1808.

The paper is structured as follows. In the beginning, we give a description of the observations of molecular spectral lines in bands 3 and 7 carried out with ALMA (section \ref{B}). In section \ref{C}, the integrated intensity distributions of the dense gas tracers and the line intensity ratios are presented, with section \ref{Cd} dedicated to the ratio of HCN (1-0) to HCO\(^{+}\) (1-0). In section \ref{D}, the physical conditions (kinetic temperature and density) of dense molecular gas in the central starburst region and outflow are estimated using a radiative transfer program. The discussion is followed by a summary of the article in section \ref{E}. The velocities in the paper are presented with respect to the local standard of rest (LSR) in radio definition.

\section{Observations}\label{B}

The ALMA observations in band 3 were designed to cover a range of frequencies (between 85.5 and 89.5 GHz) including some of the important molecular lines (Table \ref{tab1}) that arise from dense gas tracers with critical densities of the order of \(n_\mathrm{cr}\gtrsim10^4~\mathrm{cm}^{-3}\). The field imaged by the 12 m array was defined by a single pointing toward the galactic center at \((\alpha,\delta)_\mathrm{J2000}=(05^\mathrm{h}07^\mathrm{m}42\fs331\pm0\fs003,-37\degr30\arcmin45\farcs88\pm0\farcs05)\). Four basebands (center rest frequencies at 86.800, 88.716, 97.981, and 100.750 GHz, and effective bandwidths 1.875, 1.875, 0.234, and 1.875 GHz, respectively) covered the frequency range with the lines of SiO (\(v=0,~J=2-1\)), HCN (\(J=1-0\)), HCO\(^{+}\) (\(J=1-0\)), and CS (\(J=2-1)\) as the principle targets. Since all lines were observed simultaneously, the \(uv\) coverage of the visibility data is uniform across the spectrum and relative calibration uncertainties (bandpass, amplitude, and phase) are minimal. The visibility data were deconvolved and CLEANed using the Common Astronomy Software Applications (CASA) package \citep{McM07}. Two image sets were produced based on the parameters in the task ``clean'': (1) Briggs algorithm with a robustness parameter of 0.5 (optimum sensitivity and resolution) and a velocity resolution of \(\Delta v=5~\mathrm{km~s}^{-1}\) to study the small-scale structure of molecular gas distribution, and (2) natural weighting (emphasize on sensitivity) with a velocity resolution \(\Delta v=20~\mathrm{km~s}^{-1}\) to probe the extended structure. The rms sensitivity of the produced data cubes is \(\sim1\) mJy beam\(^{-1}\) (channel width \(\Delta v=5~\mathrm{km~s}^{-1}\)), and the sizes of the synthesized beams range from \(2\farcs20\times1\farcs53\) for SiO (2-1) in natural weighting to \(1\farcs15\times0\farcs75\) for CS (2-1) with robust parameter 0.5.

The observations in band 7 were focused on the lines of CO (\(J=3-2\)) and HCO\(^{+}\) (\(J=4-3\)) and 350 GHz continuum. The HCO\(^{+}\) (4-3) line was covered with one baseband at the center rest frequency of 356.734 GHz with a bandwidth of 1.875 GHz. Imaging was performed in mosaicking mode over a map of \(40\arcsec\times40\arcsec\) (blue dashed rectangle in Figure \ref{fig:n1808}). The data were acquired with the 12 m array and Atacama Compact Array (7 m antennas) and combined in the imaging (CLEAN) process. The shortest baseline of ACA was about 9 m, yielding a maximum recoverable scale of \(\sim19\arcsec\), which is comparable to the size of the central starburst region. The visibility data were deconvolved and CLEANed using CASA as for band 3. A uniform weighting image was produced for HCO\(^{+}\) (4-3) that resulted in an angular resolution of \(0\farcs98\times0\farcs52\) and an rms sensitivity of \(\sim11\) mJy beam\(^{-1}\) (channel width \(\Delta v=5~\mathrm{km~s}^{-1}\)). Throughout the article, only statistical uncertainties are given when line intensities and their ratios are calculated; the calibration uncertainty of the high signal-to-noise ratio images is of the order of \(5\%\). Since most of the lines of the intensity ratios discussed in the article (except the ratios of different CO and HCO\(^{+}\) lines) were measured simultaneously, it is assumed that the statistical uncertainties play a dominant role.

The 93 GHz continuum data from band 3 and the CO (3-2) line and 350 GHz continuum data from band 7, acquired from emission-free channels in all available basebands, were presented in \cite{Sal17} together with an observational summary that includes the observation date, on-source time, and calibrator information.

\section{Results}\label{C}

\subsection{Distribution of dense gas tracers}\label{Ca}

\begin{figure}
\epsscale{1}
\plotone{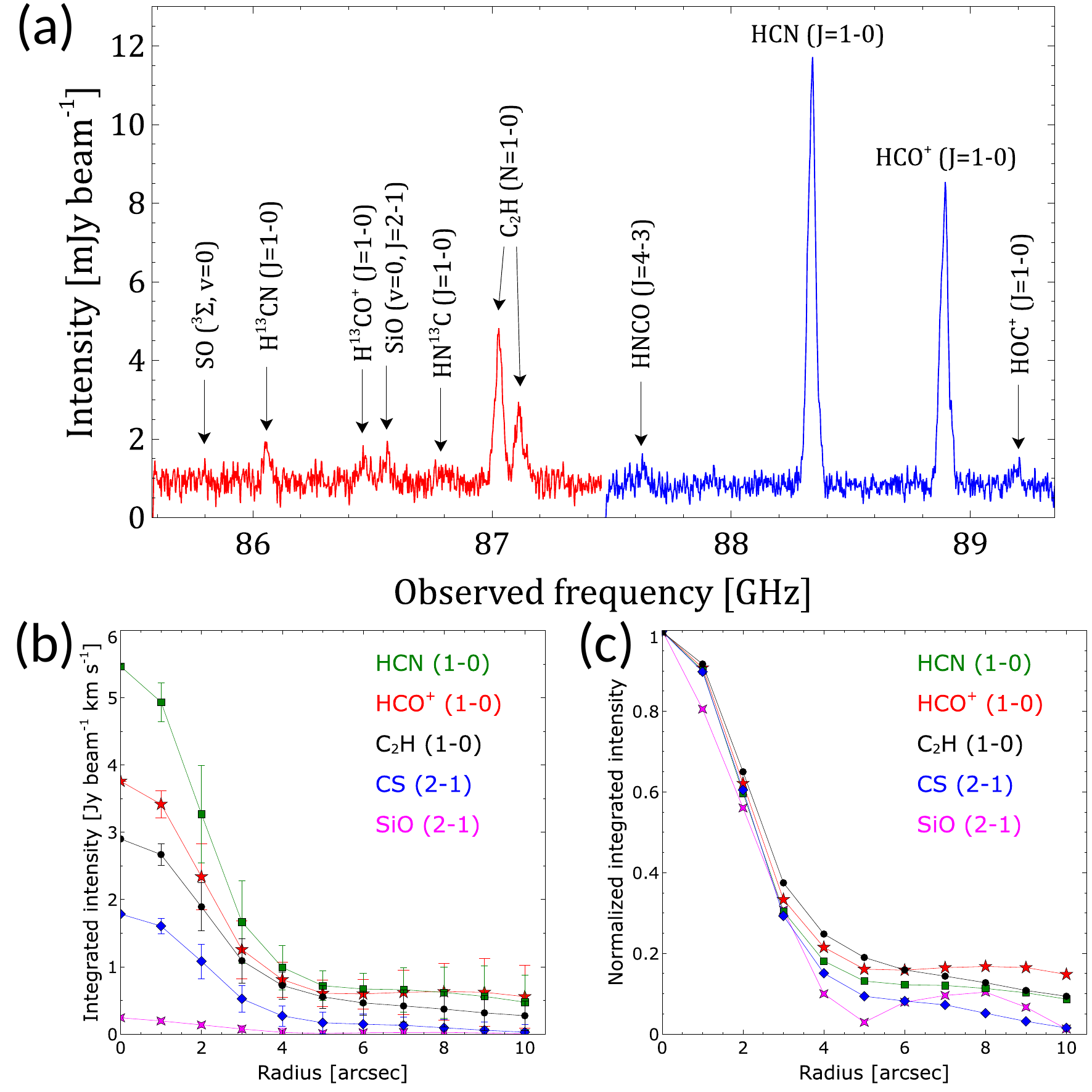}
\caption{(a) Spectrum toward the galactic center (circle with diameter \(3\arcsec\)) comprising of two adjacent basebands (lower shown in red, and upper shown in blue) covering the frequency range from 85.5 to 89.5 GHz. The spectrum was derived from the data before continuum subtraction. The spectral resolution is \(\Delta v=5~\mathrm{km~s}^{-1}\) (b) Azimuthally averaged radial profiles of the integrated intensities calculated from continuum-subtracted, natural-weighted data (resolution \(\sim2\arcsec\)) corrected for galactic position angle (324\arcdeg) and inclination (57\arcdeg). The error bars are pixel rms over the azimuthally averaged rings. The rms of integrated intensity images is \(\Delta I_\mathrm{rms}=0.1\) Jy beam\(^{-1}\) km s\(^{-1}\). The C\(_2\)H (1-0) profile was integrated over all components in Table \ref{tab1}. (c) Normalized radial profiles. The error bars are not shown in this panel for clarity.\label{fig:B3spec}}
\end{figure}

The target molecular lines were detected toward the central 1 kpc starburst disk. In this section we present their spectra and integrated intensity images, defined as \(I\equiv\int \mathcal{S} dv\), where the intensity \(\mathcal{S}\) [Jy beam\(^{-1}\)] is integrated over velocity \(v\) [km s\(^{-1}\)]. Figure \ref{fig:B3spec} shows a spectrum extracted from a circular region of diameter \(3\arcsec\) (\(\sim156\) pc) toward the galactic center, where the signal-to-noise ratio is highest for all lines. The central \(r<100\) region is referred to as the circumnuclear disk (CND). The spectrum in the figure shows a number of lines from various molecular species. The most strongly detected lines in band 3 are HCN (1-0), HCO\(^{+}\) (1-0), CS (2-1), and the doublet of C\(_2\)H (1-0), followed by the rare isotopologues H\(^{13}\)CN (1-0) and H\(^{13}\)CO\(^{+}\) (1-0), as well as SiO (2-1) and HOC\(^{+}\) (1-0). In addition, the spectrum shows emission lines of HNCO (4-3), HN\(^{13}\)C (1-0), and SO, although the detection of these is only marginal. The HCO\(^{+}\) (4-3) line was firmly detected in band 7; the line intensity ratio of HCO\(^{+}\) (4-3) to HCO\(^{+}\) (1-0) will be used as a diagnostic tool to probe the physical conditions of dense gas in section \ref{D}. All lines presented here are first detections in NGC 1808, except HCN (1-0) and HCO\(^{+}\) (1-0), which were recently measured by \cite{Gre16} at lower resolution, and HCO\(^{+}\) (4-3), which was recently imaged at high resolution toward the nucleus, although to a lower spatial extent \citep{Aud17}. Single-dish observations of HCN (1-0) toward the galactic center were also reported in \cite{Aal94}. The basic properties of the spectral lines toward the central \(3\arcsec\) aperture are listed in Table \ref{tab1}.

\begin{figure}
\epsscale{0.9}
\plotone{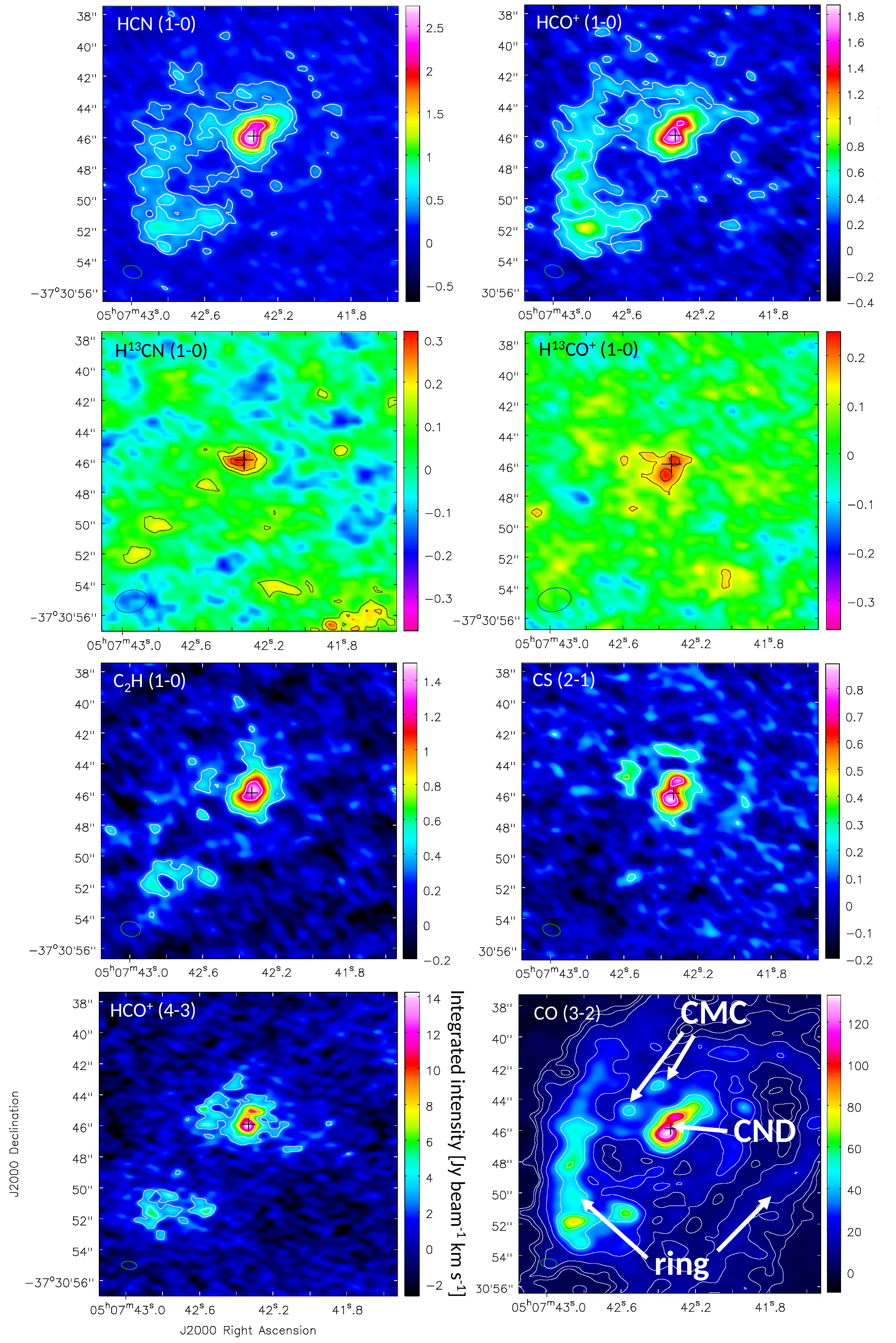}
\caption{Integrated intensity images. The contours are plotted as follows: \((3,5,10,15,20,25)\times0.1~\mathrm{Jy~beam^{-1}~km~s^{-1}}\) (\(1\sigma\)) for HCN (1-0), \((3,5,10,15)\times0.1~\mathrm{Jy~beam^{-1}~km~s^{-1}}\) for HCO\(^{+}\) (1-0), \((2,3,4)\times0.07~\mathrm{Jy~beam^{-1}~km~s^{-1}}\) for H\(^{13}\)CN (1-0),
\((2,3)\times0.07~\mathrm{Jy~beam^{-1}~km~s^{-1}}\) for H\(^{13}\)CO\(^{+}\) (1-0), \((3,5,10)\times0.12~\mathrm{Jy~beam^{-1}~km~s^{-1}}\) for C\(_2\)H (1-0), \((3,5,10)\times0.07~\mathrm{Jy~beam^{-1}~km~s^{-1}}\) for CS (2-1), \((3,5,10)\times0.9~\mathrm{Jy~beam^{-1}~km~s^{-1}}\) for HCO\(^{+}\) (4-3), and \((3,5,10,20,40,60,90,120)\times1~\mathrm{Jy~beam^{-1}~km~s^{-1}}\) for CO (3-2). The high-resolution images presented here were created using robust parameter 0.5 and velocity resolution \(\Delta v=5\) km s\(^{-1}\), except CO (3-2), H\(^{13}\)CN (1-0), and H\(^{13}\)CO\(^{+}\) (1-0), which were natural-weighted. The C\(_2\)H (1-0) image is an integrated image of all six hyperfine components in Table \ref{tab1}. The beam size is shown at the bottom left corner of each image.}\label{fig:densehr}
\end{figure}

Panels (b) and (c) of Figure \ref{fig:B3spec} show azimuthally averaged integrated intensities of the five most prominent lines derived from naturally weighted data (angular resolution \(\sim2\arcsec\)). The C\(_2\)H (1-0) profile was derived from a single integrated image of fine structure transitions (\(J=3/2-1/2\) and \(J=1/2-1/2\), each split into three hyperfine levels with \(F=J\pm1/2\)) listed in Table \ref{tab1}. All lines exhibit strong emission in the galactic center and a decrease to 0.1-0.2 of their maximum values at a radius of \(r\gtrsim5\arcsec\). The integrated intensity images (angular resolution \(\sim1\arcsec\); derived using robustness parameter 0.5) of the lines including CO (3-2) are presented in Figure \ref{fig:densehr}. At this resolution, the images show prominent emission in the CND as well as extended structure; the lines are firmly detected in the regions that surround the CND (marked by arrows in Figure \ref{fig:densehr}). Following the notation in \cite{Sal17}, these regions are referred to as the central molecular clouds (CMCs). Beyond the CMCs lies the 500 pc ring -- the outermost visible structure in the CO (3-2) integrated intensity image in the bottom right panel of Figure \ref{fig:densehr}; most lines were detected toward the southeast part of the ring.

\begin{table}
\begin{center}
\caption{Gaussian Fit Parameters of Molecular Lines Detected Toward the Central \(3\arcsec\)}\label{tab1}
\begin{tabular}{lccccccc}
\tableline\tableline
Molecule & Transition & Rest frequency & \(\mathcal{S}_\mathrm{max}\) & \(\mathcal{I}\) & \(V_\mathrm{c}\) & \(\Delta V\) \\
& & (GHz) & \((\mathrm{mJy~beam}^{-1})\) & \((\mathrm{mJy~beam^{-1}}\) & \((\mathrm{km~s^{-1}})\) & \((\mathrm{km~s^{-1}})\) \\
& & & & \(\times\mathrm{km~s^{-1}})\) & \\
\hline
SO & \(^3\Sigma,~v=0,~J_N=2_2-1_1\)\tablenotemark{(a)} & \(86.09395\) & ... & ... & ... & ... \\
H\(^{13}\)CN & \(J=1-0\)\tablenotemark{(b)} & \(86.33992\) & \(1.76\pm0.22\) & \(203\pm27\) & \(986\pm7\) & \(108\pm16\) \\
H\(^{13}\)CO\(^{+}\) & \(J=1-0\)\tablenotemark{(b)} & \(86.75429\) & \(0.91\pm0.22\) & \(143\pm40\) & \(991\pm17\) & \(145\pm45\) \\
SiO & \(v=0,~J=2-1\)\tablenotemark{(b)} & \(86.84696\) & \(1.35\pm0.20\) & \(178\pm30\) & \(988\pm9\) & \(124\pm23\) \\
HN\(^{13}\)C & \(J=1-0\)\tablenotemark{(a,b)} & \(87.09085\) & ... & ... & ... & ... \\
C\(_2\)H & \(N=1-0,~J=3/2-1/2\)\tablenotemark{(b)} & \(87.28416\)-\(87.32862\) & \(8.08\pm0.28\) & \(1155\pm45\) & \(990\pm8\)\tablenotemark{(c)} & \(134\pm6\) \\
& \((F=1-1,2-1,1-0)\) & & & & & \\
C\(_2\)H & \(N=1-0,~J=1/2-1/2\)\tablenotemark{(b)} & \(87.40200\)-\(87.44651\) & \(3.57\pm0.26\) & \(549\pm46\) & \(990\pm8\)\tablenotemark{(c)} & \(144\pm13\) \\
& \((F=1-1,0-1,1-0)\) & & & & & \\
HNCO & \(J_{Ka,Kc}=4_{0,4}-3_{0,3}\)\tablenotemark{(a,b)} & \(87.92524\) & ... & ... & ... & ... \\
HCN & \(J=1-0\) & \(88.63042\)-\(88.63393\) & \(24.5\pm0.4\) & \(3562\pm62\) & \(997.3\pm1.0\) & \(136\pm3\) \\
& \((F=1-1,2-1,0-1)\) & & & & & \\
HCO\(^{+}\) & \(J=1-0\) & \(89.18852\) & \(17.2\pm0.4\) & \(2495\pm61\) & \(999.7\pm1.4\) & \(137\pm4\) \\
HOC\(^{+}\) & \(J=1-0\)\tablenotemark{(b)} & \(89.48741\) & \(0.99\pm0.17\) & \(110\pm46\) & \(987\pm17\) & \(130\pm41\) \\
CS & \(J=2-1\)\tablenotemark{(b)} & \(97.98095\) & \(8.99\pm0.36\) & \(1157\pm51\) & \(999.4\pm2.3\) & \(121\pm6\) \\
HCO\(^{+}\) & \(J=4-3\)\tablenotemark{(b)} & \(356.73422\) & \(21.4\pm0.5\) & \(2864\pm82\) & \(1001.1\pm1.6\) & \(126\pm4\) \\
\tableline
\end{tabular}
\end{center}
\tablenotemark{(a)}{Marginal (first) detection.} \\
\tablenotemark{(b)}{First detection in NGC 1808.} \\
\tablenotemark{(c)}{The average value of the two fine-structure line sets with \(N=1-0\).}
\tablecomments{The line properties were derived from the continuum-subtracted, high-sensitivity data (natural weighting with \(\Delta v=20~\mathrm{km~s}^{-1}\); angular resolution \(\sim2\arcsec\)), after fitting a single Gaussian profile to each line. The region where the spectra were calculated is defined by a diameter of \(3\arcsec\) positioned at the galaxy center. \(V_\mathrm{c}\) is the central velocity of the Gaussian profile, and \(\Delta V\) is the FWHM of the profile. The rest frequencies are acquired from Splatalogue database for astronomical spectroscopy available at http://www.cv.nrao.edu/php/splat/.}
\end{table}

\subsection{Dense molecular gas in the CND}\label{Cb}

\begin{figure}
\epsscale{1.1}
\plotone{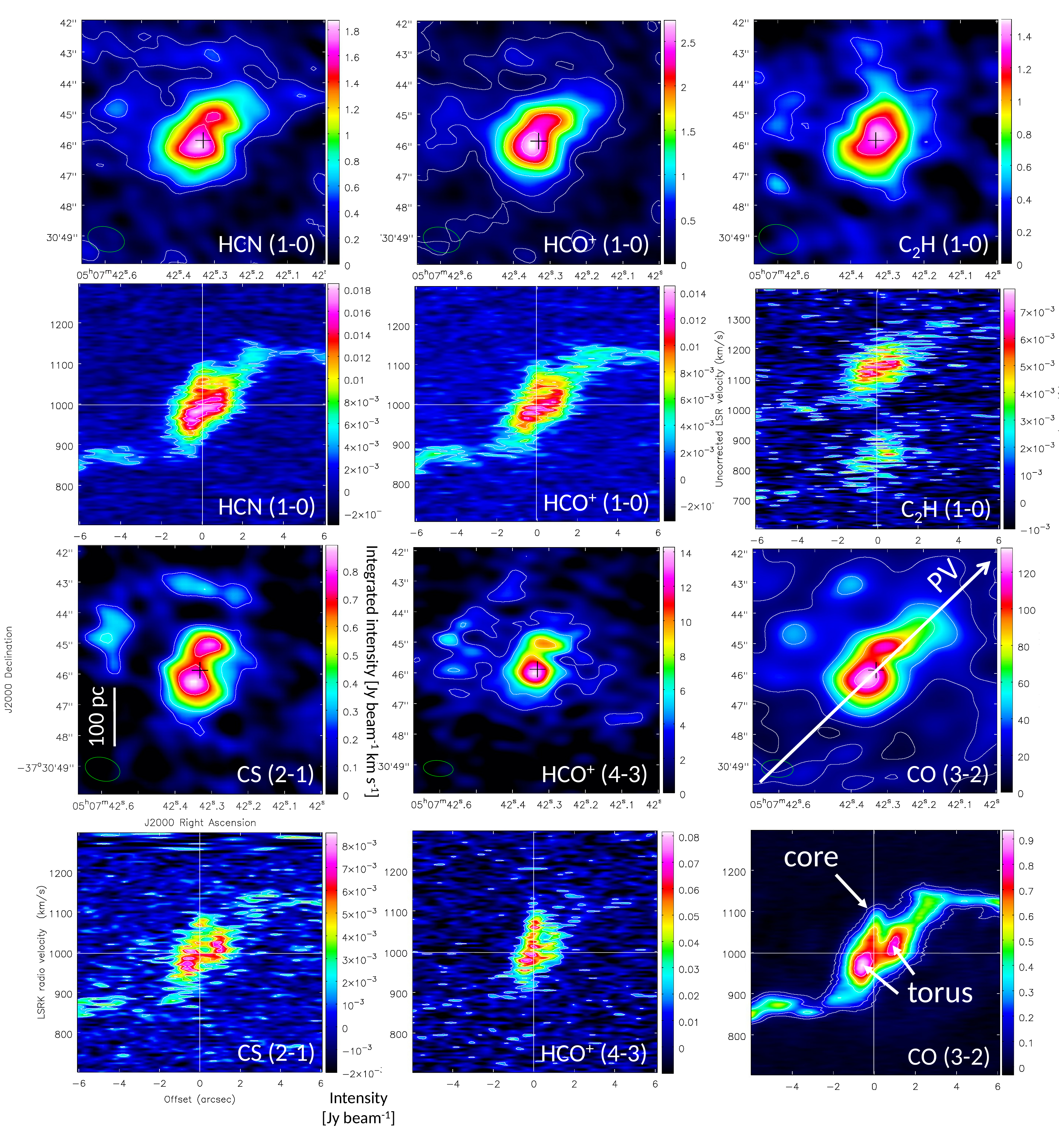}
\caption{Integrated intensities (enlarged images from Figure \ref{fig:densehr}) and position-velocity diagrams (PVDs) of dense gas tracers in the circumnuclear disk (CND). The contours in the PVDs are plotted at \((0.2, 0.4, 0.6, 0.8)\times\mathcal{I}_\mathrm{max}\), where \(\mathcal{I}_\mathrm{max}^\mathrm{HCN(1-0)}=0.0185~\mathrm{Jy~beam^{-1}}\), \(\mathcal{I}_\mathrm{max}^\mathrm{HCO^{+}(1-0)}=0.0145~\mathrm{Jy~beam^{-1}}\), \(\mathcal{I}_\mathrm{max}^\mathrm{CS(2-1)}=0.00855~\mathrm{Jy~beam^{-1}}\), and \(\mathcal{I}_\mathrm{max}^\mathrm{HCO^{+}(4-3)}=0.0817~\mathrm{Jy~beam^{-1}}\); the contours of the CO (3-2) PVD are \((0.1, 0.2, 0.4, 0.6, 0.8)\times0.937~\mathrm{Jy~beam^{-1}}\). The PVD position angle is \(\mathrm{PA}=316\arcdeg\) (major galactic axis), indicated by a white arrow denoted by ``PV''. The systemic velocity (\(V_\mathrm{sys}=998\) km s\(^{-1}\)) is marked by a horizontal white line.\label{fig:torus-dense}}
\end{figure}

Figure \ref{fig:torus-dense} shows the integrated intensity distributions of dense gas tracers in the CND. Note that the intensities of the HCN (1-0), CS (2-1), HCO\(^{+}\) (4-3), HCO\(^{+}\) (1-0), and CO (3-2) lines clearly exhibit a double peak structure. The double peak is most symmetrical in the CS (2-1) line intensity, with similar peaks offset from the galactic center (core), marked by a black plus sign. This feature has been interpreted as a molecular gas torus with a radius of \(\sim30\) pc surrounding an unresolved core offset from the midpoint between the peaks \citep{Sal17}. All tracers (except C\(_2\)H) show that the core is located closer to the brighter southeast peak of molecular gas tracers.

The C\(_2\)H (1-0) line (integrated over the fine structure components in Table \ref{tab1}) is detected in the CND with a notably different structure compared to other tracers; there is no clear indication of a double peak in the integrated intensity image of C\(_2\)H (1-0) shown in Figure \ref{fig:torus-dense}. The line has been recognized as a tracer of photodissociation regions (PDRs), dominated by strong UV radiation. PDR contain a relatively high fraction of C\(^{+}\), as well as hydrogen, resulting in efficient formation of simple molecules such as C\(_2\)H \citep{Pet05,Pet17,Mar14,Mei15,Nag15,Nis16a}. Therefore, the molecule is expected to be prominent in the gas-rich starburst nucleus of NGC 1808. The integrated intensity ratios of C\(_2\)H (1-0) to HCN (1-0) and HCO\(^{+}\) (1-0) (Table \ref{tab1}) are similar to the values obtained by \cite{Mar14} for a number of starburst and Seyfert galaxies.

In addition, HOC\(^{+}\) (1-0) was detected toward the CND. There is evidence that this molecule, an isomer of HCO\(^{+}\), is enhanced in PDRs where it can be efficiently produced, e.g., via the reactions \(\mathrm{C^{+}+H_2O\rightarrow HOC^{+}+H}\) and \(\mathrm{CO^{+}+H_2\rightarrow HOC^{+}+H}\) (e.g., \citealt{SZ94}). The abundance ratio \(\mathrm{[HCO^{+}]/[HOC^{+}]}\) in Galactic objects is usually several hundreds (e.g., \(\sim270\) in the Orion Bar and \(\sim360\) in Sgr B2 (OH); \citealt{ZA95,APT99}). In extragalactic sources, the line has been detected in the Seyfert galaxy NGC 1068 \citep{Use04} and star-forming galaxies NGC 253, M82, and M83 \citep{Ala15}, as a tracer of gas in the vicinity of star clusters with OB stars. The resolved image of HOC\(^{+}\) (1-0) in M82 suggests HOC\(^{+}\) enhancements toward regions of dense gas and PDRs \citep{Fue08}. In M82, the intensity ratio of brightness temperatures \(T_\mathrm{b,H^{13}CO^{+}(1-0)}/T_\mathrm{b,HOC^{+}(1-0)}\sim1-2\) was reported, which is comparable to the value of \(\sim1\) measured for the CND of NGC 1808 (Table \ref{tab1}). On the other hand, the detection of both HOC\(^{+}\) (1-0) and SiO (2-1) (see below) in the CND, where a hard X-ray source (possibly AGN) resides, is similar to the results obtained for NGC 1068 \citep{Use04}. The presence of these tracers allows an alternative scenario, according to which the chemistry of HOC\(^{+}\) is affected by interaction with X-rays. In both M82 and NGC 1068, the abundance of HOC\(^{+}\) seems to be enhanced compared to the Galactic objects. The line intensity (brightness temperature) ratio of HCO\(^{+}\) (1-0) to HOC\(^{+}\) (1-0) in the CND of NGC 1808 is measured to be \(19.0\pm3.8\). Although HCO\(^{+}\) (1-0) is optically thick, as shown in Appendix, the ratio is still notably low, indicating an enhanced abundance of HOC\(^{+}\).

The SiO (2-1) emission was detected toward the CND. Although relatively weak, the SiO (2-1) line is valuable because it is a well-established tracer of shocked gas (e.g., \citealt{MP92,Sch97,GB00,Kel17}). The shocks in the CND may be associated with supernova explosions and other feedback from the nuclear starburst activity exerted on the dense gas. We discuss about the distribution of SiO (2-1) integrated intensity in the context of shock chemistry in section \ref{Cd}. The tentative detection of HNCO (4-3) toward the CND is another indicator of the presence of shocked gas (e.g., \citealt{Mei15}).

In the panels of the second and forth rows, Figure \ref{fig:torus-dense} also shows position-velocity diagrams (PVDs) derived along the major galactic axis (position angle \(316\arcdeg\) determined in \citealt{Sal16}) centered at the 93 GHz continuum peak (core). The PVDs show that the kinematics of the dense gas tracers in the CND is consistent with the overall rigid body rotation of the CND and the molecular torus discussed in \cite{Sal17}. In addition, the dense gas is also present in the high velocity component denoted by the ``core'' in the lower right panel in Figure \ref{fig:torus-dense}. Note that the high-velocity core is remarkably bright in the HCO\(^{+}\) (4-3) line relative to HCO\(^{+}\) (1-0) and other tracers. Since the line is a tracer of high gas density \(n\gtrsim10^5~\mathrm{cm}^{-3}\), the data show that the core exhibits enhanced gas excitation compared to the rest of the CND. This condition will be explored in more detail in sections \ref{Cc} and \ref{D}.

\subsection{Line intensity ratios}\label{Cc}

The ALMA observations in cycles 1 and 2 have produced the data cubes of two CO lines, \(J=1-0\) and \(J=3-2\), and two HCO\(^{+}\) lines, \(J=1-0\) and \(J=4-3\), at comparable angular resolution and sensitivity (corrected for missing flux by ACA observations). The critical densities of these lines range from \(\sim10^2~\mathrm{cm}^{-3}\) for CO (1-0) to \(>10^4~\mathrm{cm}^{-3}\) for HCO\(^{+}\) (4-3) \citep{Shi15} over a wide range of temperatures found in molecular clouds, allowing us to probe the physical conditions (temperature and density) in diffuse and dense clouds across the central 1 kiloparsec (section \ref{D}). The line intensity ratio of CO (3-2) to CO (1-0), defined as \(R_\mathrm{CO}\equiv W_\mathrm{CO(3-2)}/W_\mathrm{CO(1-0)}\approx T_\mathrm{b,CO(3-2)}/T_\mathrm{b,CO(1-0)}\), where \(W=\int T_\mathrm{b} dv\) is the integrated intensity in brightness units (\(T_\mathrm{b}\) [K]), was presented in \cite{Sal17}.

\begin{figure}
\epsscale{1}
\plotone{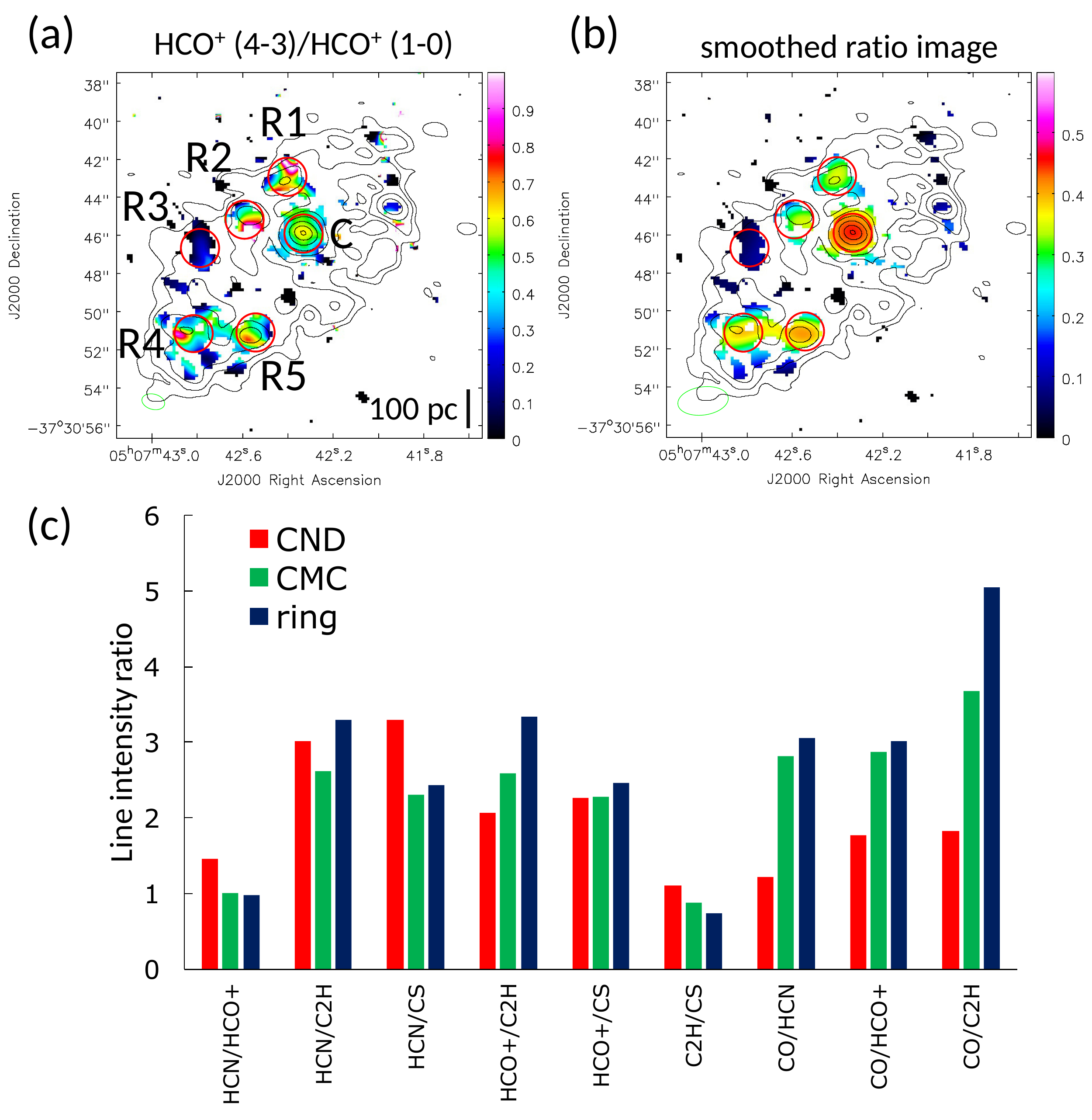}
\caption{(a) HCO\(^{+}\) (4-3) to (1-0) integrated intensity ratio image. The HCO\(^{+}\) (4-3) image was masked at \(<3~\sigma\) in intensity (where \(1~\sigma=12\) mJy beam\(^{-1}\)) prior to calculating the ratio. The contours are 93 GHz continuum plotted at \((3,5,10,20,40,70,100)\times2.5\times10^{-5}~\mathrm{Jy~beam^{-1}}\) (\(1~\sigma\)). (b) The ratio smoothed to the angular resolution of CO (1-0) for comparison. The properties of regions labeled C, R1-R5 are listed in Table \ref{tab2}. (c) Line intensity ratios in the CND (region C), CMCs (average of regions R1 and R2), and 500 pc ring (average of regions R3-R5). The CO/HCN and CO/HCO\(^{+}\) ratios are scaled by 0.2 of the original value, and the CO/C\(_2\)H ratio is scaled by 0.1.\label{fig:linrat}}
\end{figure}

The line intensity ratio of HCO\(^{+}\) (4-3) to HCO\(^{+}\) (1-0), defined as \(R_\mathrm{HCO^{+}}\equiv W_\mathrm{HCO^{+}(4-3)}/W_\mathrm{HCO^{+}(1-0)}\approx T_\mathrm{b,HCO^{+}(4-3)}/T_\mathrm{b,HCO^{+}(1-0)}\), is shown in Figure \ref{fig:linrat}. The ratio is highest in the CND: in the central \(r<1\arcsec\), the average value is \(R_\mathrm{HCO^{+}}=0.420\pm0.022\) and the maximum value within the central 50 pc is \(R_\mathrm{HCO^{+}}^\mathrm{max}(r<0\farcs5)=0.622\pm0.027\) derived at original resolution (panel (a)), where the statistical uncertainty is \(1\sigma\). Since the two lines were measured in different ALMA bands, there is also a calibration uncertainty of the order of \(5\%\); a similar error also applies to the ratio \(R_\mathrm{CO}\). The ratio is typically 0.3--0.4 in the compact sources detected around the CND. Note that the maximum value is spatially coincident with the 93 GHz continuum peak (plotted as contours) and not with the molecular torus (double peak). This result indicates that the high excitation of dense gas traced by HCO\(^{+}\) in the core may be generated by the activity of the starburst nucleus within the central \(r<30\) pc \citep{KSG94,Bus17,Sal17} possibly coexisting with an embedded Seyfert 2 type low-luminosity AGN \citep{VV85,Kot96}. The high \(R_\mathrm{HCO^{+}}\) in the core can be explained by the presence of dense gas heated by shocks.

The line ratios \(R_\mathrm{HCO^{+}}\) and \(R_\mathrm{CO}\) in Table \ref{tab2} were derived as average values within selected regions of diameter \(2\arcsec\) (circles in Figure \ref{fig:linrat}). Regions R1 and R2 correspond to the central molecular clouds (CMCs), whereas regions R3, R4, and R5 are part of the 500 pc ring. These regions have been investigated in detail because both lines of CO and HCO\(^{+}\) as well as continuum at 93 and 350 GHz were detected, allowing us to probe the gas physical conditions using a line ratio analysis. Since the region size, comparable to the size of the synthesized beam, is equivalent to a physical scale of \(\sim100\) pc, the derived values are beam-averaged and reflect mean values at 100 pc scale. The derived velocity widths \(\Delta V\) are full width half maximum (FWHM) values obtained from Gaussian fitting of the spectra within the regions.

\begin{table}
\begin{center}
\caption{Molecular Gas Parameters in Selected Regions (Diameter \(2\arcsec\))}\label{tab2}
\begin{tabular}{lccccccc}
\tableline\tableline
Region & Coordinates\tablenotemark{a} &  \(R_\mathrm{HCO^{+}}\) & \(R_\mathrm{CO}\) & \(I_\mathrm{CO(1-0)}\) & \(N_\mathrm{CO}\) & \(\Delta V_\mathrm{CO(1-0)}\) & \(\Delta V_\mathrm{HCO^{+}(1-0)}\) \\
& \((\mathrm{second},'')\) & & & \((\mathrm{Jy~km~s^{-1}})\) & \((\times10^{18}~\mathrm{cm}^{-2})\) & \((\mathrm{km~s}^{-1})\) & \((\mathrm{km~s}^{-1})\) \\
\hline
C & \(42.331,-45.877\) & \(0.420\pm0.022\) & \(0.935\pm0.025\) & 37.1 & 6.9  & \(151.1\pm2.0\) & \(135.9\pm3.8\)  \\
R1 & \(42.402,-42.909\) & \(0.303\pm0.032\) & \(0.686\pm0.063\) & 16.5 & 3.0 & \(89.0\pm3.9\) & \(65.6\pm2.4\) \\
R2 & \(42.590,-45.159\) & \(0.282\pm0.055\) & \(0.664\pm0.034\) & 23.0 & 4.2 & \(111.2\pm1.8\) & \(96.4\pm4.1\) \\
R3 & \(42.788,-46.658\) & \(0.072\pm0.020\) & \(0.641\pm0.009\) & 28.6 & 5.3 & \(97.4\pm1.7\) & \(79.9\pm4.6\) \\
R4 & \(42.815,-51.127\) & \(0.337\pm0.042\) & \(0.725\pm0.027\) & 30.0 & 5.5 & \(68.4\pm1.2\) & \(71.1\pm3.3\) \\
R5 & \(42.543,-51.068\) & \(0.356\pm0.048\) & \(0.763\pm0.042\) & 18.4 & 3.4 & \(70.2\pm1.6\) & \(53.0\pm2.1\) \\
\tableline
\end{tabular}
\end{center}
\tablenotemark{a}{The expressed coordinates have common values of \((\alpha,\delta)=(05^\mathrm{h}07^\mathrm{m},-37\degr30\arcmin)\).}
\tablecomments{The \(R_\mathrm{HCO^{+}}\) and \(R_\mathrm{CO}\) ratios were derived at a common angular resolution of the CO (1-0) line (\(2\farcs670\times1\farcs480\)) from \cite{Sal17}. The uncertainty is \(1~\sigma\) of the mean value; it does not include the calibration (absolute intensity) uncertainty of \(\sim5\%\).}
\end{table}

In Table \ref{tab3}, we list the intensity ratios of the band 3 lines HCN (1-0), HCO\(^{+}\) (1-0), C\(_2\)H (1-0), CS (2-1), and CO (1-0) detected throughout regions C and R1-R5 defined in Figure \ref{fig:densehr}. In general, HCN (1-0) and CS (2-1) are regarded as dense gas tracers, HCO\(^{+}\) (1-0) and C\(_2\)H (1-0) are expected to be bright in PDRs, whereas CO (1-0) is a tracer of bulk molecular gas. High-sensitivity data cubes (natural weighting) were used in deriving the line intensity ratios. Below, we summarize some clear trends, observed in the line ratios, reflecting the complex physical and chemical structure of the starburst region. The trends are illustrated in panel (c) of Figure \ref{fig:linrat}.

\begin{enumerate}
\item{All dense gas tracers, such as HCN (1-0), HCO\(^{+}\) (1-0), and C\(_2\)H (1-0), are generally enhanced with respect to CO (1-0) in the CND (region C).}
\item{HCN (1-0) is enhanced with respect to HCO\(^{+}\) (1-0) and CS (2-1) in the CND. The ratios of HCN (1-0) with respect to these two lines are remarkably uniform among regions R1-R5. In the CND, the ratios of HCN (1-0) with respect to HCO\(^{+}\) (1-0) and CS (2-1) are \(\sim1.5\) and \(\sim1.4\) larger compared to average of the ratios in the other regions, respectively.}
\item{The ratio of HCN (1-0) to HCO\(^{+}\) (1-0) is close to unity in all regions except in the CND.}
\item{C\(_2\)H (1-0) is slightly enhanced with respect to HCO\(^{+}\) (1-0) and CS (2-1) in the CND, showing line intensity ratios similar to those of HCN (1-0).}
\item{C\(_2\)H (1-0) intensity is generally low compared to HCN (1-0), HCO\(^{+}\) (1-0), and CS (2-1) in the 500 pc ring (regions R3-R5).}
\item{The ratio of HCO\(^{+}\) (1-0) to CS (2-1) exhibits no significant variation among all investigated regions.}
\end{enumerate}

The presence of strong emission from dense gas tracers HCN (1-0), HCO\(^{+}\) (1-0), CS (2-1), and C\(_2\)H (1-0) suggests that the CND is a large PDR, abundant in dense gas and dominated by star formation feedback. The detection of tracers such as SiO (2-1) toward the nucleus support the picture of shocked dense gas. Below, we present the ratio of HCN (1-0) to HCO\(^{+}\) (1-0) in more detail and discuss on its possible origin.

\begin{table}
\begin{center}
\caption{Line Intensity Ratios in Selected Regions (Diameter \(2\arcsec\))}\label{tab3}
\begin{tabular}{lccccccccc}
\tableline\tableline
Region & \(\frac{\mathrm{HCN(1-0)}}{\mathrm{HCO^{+}(1-0)}}\) & \(\frac{\mathrm{HCN(1-0)}}{\mathrm{C_2H(1-0)}}\) & \(\frac{\mathrm{HCN(1-0)}}{\mathrm{CS(2-1)}}\) & \(\frac{\mathrm{HCO^{+}(1-0)}}{\mathrm{C_2H(1-0)}}\) & \(\frac{\mathrm{HCO^{+}(1-0)}}{\mathrm{CS(2-1)}}\) & \(\frac{\mathrm{C_2H(1-0)}}{\mathrm{CS(2-1)}}\) & \(\frac{\mathrm{CO(1-0)}}{\mathrm{HCN(1-0)}}\) & \(\frac{\mathrm{CO(1-0)}}{\mathrm{HCO^{+}(1-0)}}\) & \(\frac{\mathrm{CO(1-0)}}{\mathrm{C_2H(1-0)}}\) \\
\hline
C & \(1.46\pm0.04\) & \(3.01\pm0.12\) & \(3.30\pm0.11\) & \(2.06\pm0.09\) & \(2.26\pm0.08\) & \(1.10\pm0.04\) & \(6.09\pm0.20\) & \(8.88\pm0.38\) & \(18.3\pm1.2\) \\
R1 & \(0.98\pm0.05\) & \(2.65\pm0.26\) & \(2.12\pm0.12\) & \(2.69\pm0.25\) & \(2.15\pm0.12\) & \(0.80\pm0.07\) & \(11.6\pm1.1\) & \(11.5\pm0.9\) & \(30.8\pm5.0\) \\
R2 & \(1.04\pm0.05\) & \(2.59\pm0.31\) & \(2.50\pm0.18\) & \(2.49\pm0.29\) & \(2.41\pm0.17\) & \(0.97\pm0.10\) & \(16.6\pm1.2\) & \(17.2\pm1.1\) & \(42.8\pm8.1\) \\
R3 & \(1.00\pm0.06\) & \(3.68\pm0.34\) & \(2.90\pm0.26\) & \(3.66\pm0.36\) & \(2.89\pm0.28\) & \(0.80\pm0.08\) & \(18.1\pm1.2\) & \(18.2\pm1.5\) & \(66.5\pm9.7\) \\
R4 & \(0.96\pm0.05\) & \(2.87\pm0.23\) & \(2.35\pm0.19\) & \(2.98\pm0.24\) & \(2.44\pm0.20\) & \(0.82\pm0.07\) & \(16.0\pm1.1\) & \(15.4\pm1.1\) & \(46.1\pm5.5\) \\
R5 & \(0.99\pm0.04\) & \(3.34\pm0.44\) & \(2.04\pm0.13\) & \(3.38\pm0.45\) & \(2.06\pm0.13\) & \(0.61\pm0.07\) & \(11.7\pm0.7\) & \(11.5\pm0.8\) & \(39.0\pm8.5\) \\
\tableline
\end{tabular}
\end{center}
\tablecomments{The ratios of intensities (in units K) are derived from Gaussian fitting of spectral lines. The C\(_2\)H (1-0) line refers to the triplet \(J=3/2\rightarrow1/2\). The uncertainties include statistical errors, but not calibration (absolute intensity) uncertainties (\(\sim5\%\) for high signal-to-noise ratio).}
\end{table}

\subsection{HCN (1-0) to HCO\(^{+}\) (1-0) line intensity ratio and shocked gas}\label{Cd}

The most firmly detected dense gas tracers in band 3 are HCN (1-0) and HCO\(^{+}\) (1-0). Since the discovery of an enhancement of the intensity ratio of the two lines, defined as \(R_\mathrm{H}\equiv W_\mathrm{HCN(1-0)}/W_\mathrm{HCO^{+}(1-0)}\approx T_\mathrm{b,HCN(1-0)}/T_\mathrm{b,HCO^{+}(1-0)}\), where \(W\equiv \int T_\mathrm{b}dv\) is the integrated intensity, in a number of active galactic nuclei, the ratio has been used as an indicator of Seyfert activity in galaxies \citep{Koh01,Koh03,INK06}. The origin of high \(R_\mathrm{H}\), also observed for higher \(J\) transitions \citep{Hsi12,Izu13,Izu16,IN14}, is not fully understood, but some authors have suggested the following explanations: difference in gas density (the critical density of HCN (1-0) is larger than that of HCO\(^{+}\) (1-0)), high-temperature chemistry (radiative or mechanical heating), X-ray dominated chemistry (in X-ray dominated regions; XDRs), infrared pumping, and electron excitation (e.g., \citealt{MSI06,Kri08,HHW10,HTH13,Taf10,Aal12,Aal15,Izu13,Mat15,GK17}). A difference in opacities may also be a cause of variation in optically thick nuclear regions \citep{MT12,MTB14}. If the two lines are moderately optically thick, and we have estimated that \(\tau>1\) in the core of NGC 1808 (Appendix), then the diagnostic method needs to be supported by independently measured column densities of the two species, to clarify whether the ratio reflects the actual abundance ratio. We performed this calculation (under LTE approximation) in Appendix for the center position and found that \(N_\mathrm{HCN}>N_\mathrm{HCO^{+}}\) in the CND region. This result is also supported by the line intensity ratio of the \(^{13}\)C isotopic species of the two molecules, which is found to be \(\mathcal{S}_{\mathrm{H^{13}CN}(1-0)}/\mathcal{S}_{\mathrm{H^{13}CO^{+}}(1-0)}=1.93\pm0.53\). If the two lines are optically thin, the ratio is proportional to the ratio of column densities as \(\mathcal{S}_{\mathrm{H^{13}CN}(1-0)}/\mathcal{S}_{\mathrm{H^{13}CO^{+}}(1-0)}\approx T_\mathrm{b,H^{13}CN(1-0)}/T_\mathrm{b,H^{13}CO^{+}(1-0)}=(\mu_\mathrm{H^{13}CN}/\mu_\mathrm{H^{13}CO^{+}})^2 N_\mathrm{H^{13}CN}/N_\mathrm{H^{13}CO^{+}}\), where the ratio of permanent dipole moments of the two molecules is \((\mu_\mathrm{H^{13}CN}/\mu_\mathrm{H^{13}CO^{+}})^2=0.59\) (e.g., \citealt{MS15}). From these considerations, the column density ratio of the two isotopic species becomes \(N_\mathrm{H^{13}CN}/N_\mathrm{H^{13}CO^{+}}=3.3\pm0.9\).

\begin{figure}
\epsscale{1}
\plotone{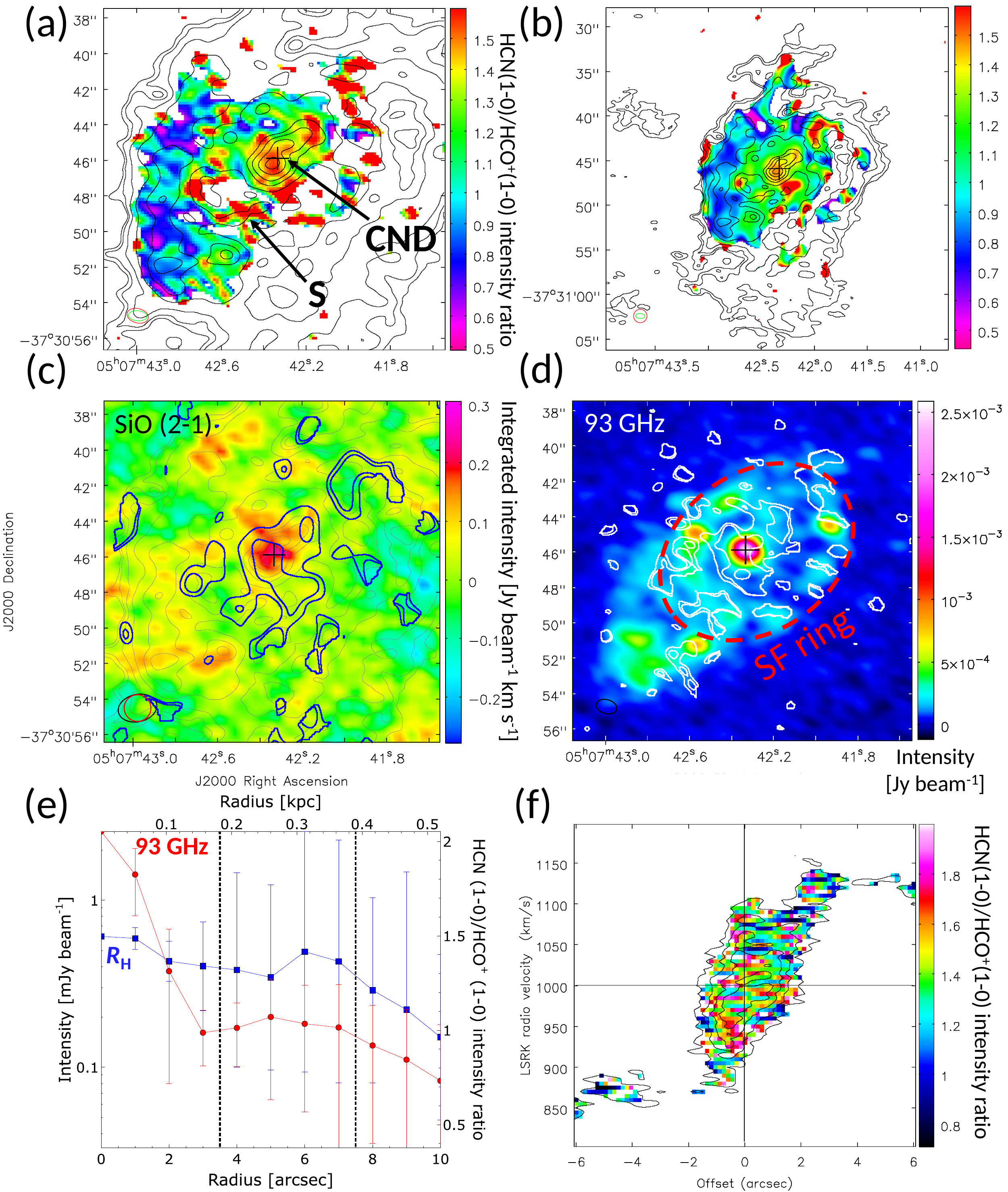}
\caption{(a) HCN (1-0) to HCO\(^{+}\) (1-0) integrated intensity ratio \(R_\mathrm{H}\) at original resolution with CO (3-2) contours as in Figure \ref{fig:densehr}. The ratio is masked below \(2~\sigma\) of integrated intensity images. The regions with enhanced ratio are indicated: CND and the nuclear spiral arm (S). (b) The ratio derived from images smoothed to an angular resolution of \(1\farcs5\times1\farcs5\). The beam size is shown at the bottom left corner. (c) SiO (2-1) integrated intensity image with CO (3-2) intensity contours (gray) and \(R_\mathrm{H}=(1.2,1.4)\) contours (blue) at resolution \(1\farcs5\) as in panel (b). (d) 93 GHz continuum image with \(R_\mathrm{H}=(1.2,1.4)\) contours (white) at resolution \(\sim1\arcsec\) as in panel (a). A star-forming (SF) ring is indicated with a dashed ellipse. (e) Azimuthally averaged radial profile of the ratio in panel (b) along with the 93 GHz continuum intensity (left vertical axis). The dashed vertical lines mark the spatial extent of the SF ring. The error bars are \(1\sigma\). (f) The line intensity ratio in position-velocity space along galactic major axis with HCN (1-0) contours as in Figure \ref{fig:torus-dense}.\label{fig:ratios}}
\end{figure}

The distributions of the line intensity ratios, \(R_\mathrm{H}\), across the central 1 kpc starburst region of NGC 1808, are shown in Figure \ref{fig:ratios}. The figure reveals the following important characteristics of the measured ratios: (1) \(R_\mathrm{H}\) is enhanced in the CND (central region of 200 pc diameter) where the pixel-averaged ratio is \(R_\mathrm{H}(r<2\arcsec)=1.45\pm0.08\); (2) \(R_\mathrm{H}\) is also enhanced to \(\sim1.5\) in a narrow, elongated feature indicated with an arrow in Figure \ref{fig:ratios}(a). The ``S'' feature coincides with a molecular spiral pattern distributed around the CND \citep{Sal17}. Figure \ref{fig:ratios}(b) shows a spatially smoothed distribution of \(R_\mathrm{H}\): the ratio is at its maximum in the galactic center and decreases to near unity throughout the starburst disk (see Table \ref{tab3}). This result is also apparent in panel (e), where azimuthally-averaged \(R_\mathrm{H}\) is plotted as a function of galactocentric distance. Note also in panel (f) that there is no indication of significant enhancement of \(R_\mathrm{H}\) in the core (high velocity component at offset \(0\arcsec\) and \(1050~\mathrm{km~s}^{-1}\lesssim V_\mathrm{LSR}\lesssim1100~\mathrm{km~s}^{-1}\)) compared to the rest of the CND. This behavior is different from the excitation of HCO\(^{+}\), expressed as a ratio of brightness temperatures \(R_\mathrm{HCO^{+}}\equiv T_\mathrm{b,HCO^{+}(4-3)}/T_\mathrm{b,HCO^{+}(1-0)}\), which is clearly elevated in the core (shown below in Figure \ref{fig:pvdr}(b)), and implies that caution is needed when \(R_\mathrm{H}\) is interpreted as a result of gas excitation.

Applying the nuclear activity diagnostic method of \cite{Koh01} and \cite{INK06}, the nucleus of NGC 1808, with its ratios of \(T_\mathrm{b,HCN(1-0)}/T_\mathrm{b,HCO^{+}(1-0)}=1.46\pm0.04\) and \(T_\mathrm{b,HCN(1-0)}/T_\mathrm{b,CO(1-0)}=0.164\pm0.005\), lies in the boundary region between pure AGN and starburst activity, and thus can be regarded as a composite of coexisting AGN and starburst. This result is in agreement with the observed activity of the nucleus, where starburst activity and a tentative AGN have been found.

The origin of the distribution of \(R_\mathrm{H}\) in the CND and spiral pattern is not clear. Similar results have been reported, e.g., from ALMA observations of the CNDs of the nearby Seyfert galaxies NGC 1068, NGC 1097, and NGC 613 by \cite{GB14}, \cite{Mar15}, and \cite{Miy17}, respectively. Although NGC 1808 is not a ``pure'' AGN, its nuclear value of \(R_\mathrm{H}\) is still enhanced with respect to the surrounding starburst regions, the observation that requires explanation. One recently proposed scenario is that \(R_\mathrm{H}\) can be enhanced in the CNDs of Seyfert galaxies in the environment of high temperature gas chemistry \citep{HHW10,HTH13}. Indeed, high gas temperatures have been measured in the nuclei of Seyfert \citep{Izu13,Miy15} and starburst galaxies \citep{Gor17,And17}. Then the question is: what is the heating source in starburst-dominated galactic nuclei? With \(L_\mathrm{X}\sim1\times10^{39}~\mathrm{erg~s}^{-1}\), the X-ray luminosity of the AGN in NGC 1808 is relatively weak \citep{JB05,Ter02}, and there is no evidence of a plasma jet from the core. Furthermore, the spiral arm that exhibits high \(R_\mathrm{H}\) is \(\sim200\) pc away from the galactic core, making the radiation from the AGN an unlikely driving force. In this context, we consider the possibility of star formation feedback. In a vigorous starburst, multiple supernova explosions are the main sources of interstellar shocks and a flux of cosmic rays. A similar explanation that involves shocks has been proposed for the observed line intensities in the outflow of Mrk 231 \citep{Aal12}. This scenario seems plausible for the CND and nuclear spirals in NGC 1808, because shocks produced by cloud collisions, supernova explosions, and other feedback in H II regions can heat the gas. Since numerous ``hot spots'' are found throughout the central 1 kpc disk, elevated \(R_\mathrm{H}\) in principle should not be confined to the galactic center.

In order to explore the scenario of a chemistry supported by shocks, we investigated the intensity distribution of SiO (2-1) and the 93 GHz continuum shown in panels (c) and (d) of Figure \ref{fig:ratios}. While SiO (2-1) is a well-known tracer of dense shocked gas, the 93 GHz continuum is dominated by free-free thermal emission produced in H II regions and a minor contribution of synchrotron nonthermal emission from supernova remnants \citep{Sal17}. This is where molecular clouds are in direct contact with hot gas (PDRs) ionized by massive stars and shocks in supernova explosions. Since SiO (2-1) is detected only toward the nucleus (Figure \ref{fig:ratios}(c)), it does not reveal any clear spatial correlation with the enhanced \(R_\mathrm{H}\) beyond the CND. In the CND, however, the intensity peak of SiO (2-1) is coincident with \(R_\mathrm{H}\) at the resolution of \(2\arcsec\). Furthermore, there is a marginal spatial correlation between the 93 GHz continuum and high \(R_\mathrm{H}\) even beyond the CND; the continuum exhibits a patchy ring at a radius of \(\sim300\) pc (Figure \ref{fig:ratios}(d)) and coincides with the distributions of star formation tracers such as Pa\(\alpha\), [Fe II], and hot hydrogen gas \citep{Bus17}. Note that this ring is smaller than the 500 pc ring marked in Figure \ref{fig:densehr}. Figure \ref{fig:ratios}(e) shows an increase in the azimuthally averaged \(R_\mathrm{H}\) at \(r\sim6\arcsec\); this radius corresponds to the star-forming ring.  Although modest values of \(R_\mathrm{H}\) are found at the continuum peaks along the ring, many of the high \(R_\mathrm{H}\) regions lie near the peaks inside the ring. This may imply that the ratio \(R_\mathrm{H}\) in the nuclear spiral pattern, which is a part of the ring, is in some way affected by star formation activity.

The detections of SiO, HNCO (Figure \ref{fig:B3spec}) and radio continuum indicate the presence of shocked gas in the CND and CMCs. Although shocked gas is a favorable environment for the enhancement of \(R_\mathrm{H}\) in the high temperature chemistry scenario, further observations of SiO or CH\(_3\)OH, which is another well-established shock tracer, at higher sensitivity and resolution would clarify whether the observed \(R_\mathrm{H}\) is closely related to shock heating. Also, the possibility of electron excitation \citep{GK17} as a key factor in regulating \(R_\mathrm{H}\) in regions such as the SF ring cannot be ruled out, especially if the ratio is predominantly large in cloud envelopes where gas densities are relatively low \citep{Kau17}. The observed correlation of \(R_\mathrm{H}\) with the 93 GHz continuum implies that electron excitation may be responsible for the observed ratios to some extent. Higher angular resolution observations of the SF ring and CND could shed light on the issue. It should be noted that the line intensity ratios involving C\(_2\)H (1-0), which is a PDR tracer, behave in a similar way as those of HCN (1-0) with respect to tracers such as HCO\(^{+}\) (1-0), CS (2-1), and CO (1-0) (Figure \ref{fig:linrat}(c)); the C\(_2\)H (1-0) line, which is moderately optically thin (see Appendix), is prominent in the CND, indicating high column densities. These trends can be tested in chemical models based on star formation feedback (PDR environment). To analyze the observed line intensity ratios in greater detail, it is important to estimate the physical conditions in the CND and the rest of the starburst disk, which is the subject of the next section.

\section{Physical conditions of molecular gas}\label{D}

The measured intensities of various spectral lines in bands 3 and 7 can be utilized to investigate the physical conditions of molecular gas in the starburst region. In this section, we present non-LTE (local thermodynamic equilibrium) radiative transfer calculations that yield the gas density and temperature in five distinct regions. The regions are defined as follows (see Figures \ref{fig:linrat} and \ref{fig:radex}(a)): (1) CND (region C in Table \ref{tab2}), (2) CMC (average of regions R1 and R2), (3) downstream side of the 500 pc ring (regions R4 and R5), (4) upstream side of the ring (region R3), and (5) the nuclear gas outflow (denoted by the ``stream'' in the figures below). The regions C and R1-R5 in the starburst disk represent the massive (\(\sim10^6-10^7~M_\sun\)) molecular clouds discussed in \cite{Sal17}. The analysis presented below is therefore focused on the compact sources revealed in multi-line observations; a more general analysis of the starburst disk that includes the diffuse inter-cloud medium is beyond the scope of this article. We also investigated the physical conditions of the core within the CND. The core is defined as the high-velocity component in the position-velocity space of CO and HCO\(^{+}\) (Figures \ref{fig:torus-dense} and \ref{fig:pvdr}). Spatially, it is coincident with the location of the continuum peak (Figure \ref{fig:torus-dense}).

An LTE analysis of the optical depths and column densities of dense gas tracers HCN, HCO\(^{+}\), and C\(_2\)H toward the CND is presented in Appendix. The LTE calculations were possible for the central region owing to the detection of the rare isotopic species H\(^{13}\)CN (1-0) and H\(^{13}\)CO\(^{+}\) (1-0), and well-resolved fine structure of C\(_2\)H (1-0).

\subsection{Non-LTE calculation of gas density and kinetic temperature}\label{Da}

The non-LTE calculations were carried out using the radiative transfer program RADEX \citep{vdT07} applied on the line intensity ratios of CO (3-2)/CO (1-0), HCO\(^{+}\) (4-3)/HCO\(^{+}\) (1-0), HCN (1-0)/H\(^{13}\)CN (1-0), HCO\(^{+}\) (1-0)/H\(^{13}\)CO\(^{+}\) (1-0), HCN (1-0)/HCO\(^{+}\) (1-0) and H\(^{13}\)CN (1-0)/H\(^{13}\)CO\(^{+}\) (1-0). The ratios that involve the \(^{13}\)C-bearing molecules are available only for the CND where H\(^{13}\)CN (1-0) and H\(^{13}\)CO\(^{+}\) (1-0) were firmly detected. Nevertheless, the analysis of the CND yields important information regarding the physical conditions of the dense gas, because all four \(J=1-0\) lines of hydrogen cyanide and formylium have comparable critical densities and, unlike CO (1-0), can be regarded as probes of gas with densities of the order \(n_\mathrm{H_2}\gtrsim10^3~\mathrm{cm}^{-3}\). Another advantage of using the \(^{13}\)C-bearing molecules is that their transitions are likely optically thin; the ratios of column densities, necessary in RADEX calculations, can be constrained. For a given molecular column density and a 1-dimensional velocity width, \(\Delta V\), the program can derive the line intensity ratio as a function of kinetic temperature (\(T_\mathrm{k}\)) and density of molecular (hydrogen) gas (\(n_\mathrm{H_2}\)). We assume a background temperature of \(T_\mathrm{bg}=2.73~\mathrm{K}\) due to the cosmic microwave background radiation. The geometry related to the photon escape probability is set to be an expanding sphere, equivalent to the large velocity gradient approximation \citep{Sob57,SS74,GW74,GYL83}. The ratio of the column density of a molecular species X to the velocity width, \(N_\mathrm{X}/\Delta V\) is related to a velocity gradient \(\Delta V/\Delta r\) via \(N_\mathrm{X}=f_\mathrm{X}n_\mathrm{H_2}\Delta V/(\Delta V/\Delta r)\), where \(f_\mathrm{X}\equiv[\mathrm{X}]/[\mathrm{H}_2]\) is the relative abundance, and \(n_\mathrm{H_2}\) is the number density of H\(_2\) molecules.

The column density \(N_\mathrm{CO}\) is estimated from the CO (1-0) integrated intensity (within \(2\arcsec\) regions) applying a CO-to-H\(_2\) conversion factor of \(X_\mathrm{CO}=0.8\times10^{20}~\mathrm{cm}^{-2}(\mathrm{K~km~s}^{-1})^{-1}\) \citep{Sal14} and an abundance ratio of \([\mathrm{CO}]/[\mathrm{H_2}]=10^{-4}\). The equation applied is \(N_\mathrm{H_2}=X_\mathrm{CO}W_\mathrm{CO}\), where \(W_\mathrm{CO}\) is the integrated intensity in K km s\(^{-1}\). The resulting column densities of H\(_2\) are of the order \(N_\mathrm{H_2}\sim3\mathrm{-}7\times10^{22}~\mathrm{cm}^{-2}\) (Table \ref{tab2}). A Galactic conversion factor of \(X_\mathrm{CO}=2\times10^{20}~\mathrm{cm}^{-2}(\mathrm{K~km~s}^{-1})^{-1}\) would yield column densities of \(N_\mathrm{H_2}\sim1\mathrm{-}2\times10^{23}~\mathrm{cm}^{-2}\) averaged over 100 pc. In general, the conversion factor is applicable in the case of optically thick molecular clouds in virial equilibrium, and therefore results in an upper limit for the molecular gas mass \citep{Sol87}. Although CO (1-0) emission is usually optically thick in molecular clouds in galactic disks, the optical depth in a starburst nucleus can have moderate values due to high velocity dispersion (e.g., \(\tau_\mathrm{CO(1-0)}\sim2\mathrm{-}5\) in NGC 253; \citealt{Mei15}).

A lower limit of \(N_\mathrm{CO}\) toward the galactic center position can be obtained by assuming optically thin CO (1-0) emission in LTE. For an abundance ratio \([\mathrm{CO}]/[\mathrm{H_2}]=10^{-4}\) and excitation temperature \(T_\mathrm{ex}=31\) K, we obtain \(N_\mathrm{H_2}\sim1\times10^{22}~\mathrm{cm}^{-2}\), expectedly smaller than the value derived by using \(X_\mathrm{CO}\). Note that a lower excitation temperature would result in a lower column density.

The column density of \(N_\mathrm{HCO^{+}}\) is more difficult to estimate, because \(\tau_\mathrm{HCO^{+}(1-0)}>1\). In this analysis, we adopt an abundance ratio of \([\mathrm{HCO^{+}}]/[\mathrm{H_2}]=10^{-8}\) reported in previous works (e.g., \citealt{Hog97}). This is a crude estimate, because the abundance of HCO\(^{+}\), as a molecular ion, is highly sensitive to the ionization degree of molecular gas, that may vary between regions and decrease in dense gas due to recombination (e.g., \citealt{Pap07}). This assumption yields the column densities of the order \(N_\mathrm{HCO^{+}}\sim10^{14}~\mathrm{cm}^{-2}\), consistent with those observed in dark molecular clouds in the Galaxy \citep{San12} and nearby star-forming galaxies \citep{Ala15}.

The FWHM velocity width (\(\Delta V\)) can be calculated from Gaussian fitting of the spectra toward the selected regions. Averaged over 100 pc, the velocity widths range between 50 and 150 km s\(^{-1}\) and are typically somewhat larger for CO (1-0) than for HCO\(^{+}\) (1-0) in most of the regions (Table \ref{tab2}). Some of the line broadening in the central 100 pc region is produced by galactic rotation. Extremely large line widths in regions R1-R5 may be an indicator of small-scale dynamics (e.g., cloud rotation and relative velocities of unresolved clouds) and stellar feedback (e.g., gas outflows). All calculations were conducted using \(\Delta V=50~\mathrm{km~s}^{-1}\), which is a typical observed value within individual pixels of the image. Since the results of RADEX calculations depend on the ratio \(N/\Delta V\), we fix \(\Delta V\) and vary \(N\) over three orders of magnitude (section \ref{Db}).

The results of calculations are shown in Figures \ref{fig:radex0}, \ref{fig:radex}, and summarized in Table \ref{tab4}. The plotted curves are the average values of the observed line intensity ratios in the selected \(2\arcsec\) regions (Table \ref{tab2}). The physical conditions (density and temperature) are estimated from the areas where the intensity ratios intersect in the parameter space. Below, we discuss some important characteristics of the diagrams for each investigated region. The analysis begins with the dense gas tracers in the CND and then expands to other circumnuclear regions.

\subsection{Dense gas conditions in the CND}\label{Daa}

The dense gas in the CND is first analyzed by using the line intensities of HCN (1-0), H\(^{13}\)CN (1-0), HCO\(^{+}\) (1-0), HCO\(^{+}\) (4-3), and H\(^{13}\)CO\(^{+}\) (1-0), whose ratios are defined as \(R_\mathrm{HCO^{+}}\equiv T_\mathrm{b,HCO^{+}(4-3)}/T_\mathrm{b,HCO^{+}(1-0)}=0.420\pm0.022\), \(R_\mathrm{H}\equiv T_\mathrm{b,HCN(1-0)}/T_\mathrm{b,HCO^{+}(1-0)}=1.46\pm0.04\), \(r_\mathrm{HCN}\equiv T_\mathrm{b,HCN(1-0)}/T_\mathrm{b,H^{13}CN(1-0)}=12.8\pm1.8\), \(r_\mathrm{HCO^{+}}\equiv T_\mathrm{b,HCO^{+}(1-0)}/T_\mathrm{b,H^{13}CO^{+}(1-0)}=17.0\pm7.1\), and \(r_\mathrm{H}\equiv T_\mathrm{b,H^{13}CN(1-0)}/T_\mathrm{b,H^{13}CO^{+}(1-0)}=1.93\pm0.53\). The RADEX calculations were carried out in a temperature range of \(5~\mathrm{K}\leq T_\mathrm{k}\leq500~\mathrm{K}\) and a density range of \(10^2~\mathrm{cm^{-3}}\leq n_\mathrm{H_2}<10^7\leq\mathrm{cm^{-3}}\), where the applied molecular parameters were acquired from the Leiden Atomic and Molecular Database\footnote{http://home.strw.leidenuniv.nl/\(\sim\)moldata/} \citep{Sch05}. We assumed that \([\mathrm{HCO^{+}}]/[\mathrm{H}_2]=10^{-8}\) and \([^{12}\mathrm{C}]/[^{13}\mathrm{C}]\sim20\mathrm{-}50\) \citep{Sal14}, and the velocity width was fixed to \(\Delta V=50\) km s\(^{-1}\). The calculations were carried out for four setups, as described below.

In setup 1, the applied column densities were: \(N_\mathrm{HCN}=2.3\times10^{15}~\mathrm{cm}^{-2}\), \(N_\mathrm{H^{13}CN}=6.6\times10^{13}~\mathrm{cm}^{-2}\), \(N_\mathrm{HCO^{+}}=7.0\times10^{14}~\mathrm{cm}^{-2}\), and \(N_\mathrm{H^{13}CO^{+}}=2.0\times10^{13}~\mathrm{cm}^{-2}\). The values were selected to satisfy \([^{12}\mathrm{C}]/[^{13}\mathrm{C}]=35\), \(N_\mathrm{H^{13}CN}/N_\mathrm{H^{13}CO^{+}}=3.3\), derived assuming optically thin emission of these isotopic species (section \ref{Cd}), and the same HCN to HCO\(^{+}\) abundance ratio; note that this ratio assumes that the abundance of HCN is not significantly enhanced with respect to HCO\(^{+}\), and that the emission is close to LTE.

Setup 2 adopted the same column density of HCO\(^{+}\), but a higher ratio \([^{12}\mathrm{C}]/[^{13}\mathrm{C}]=50\), resulting in \(N_\mathrm{H^{13}CO^{+}}=1.4\times10^{13}~\mathrm{cm}^{-2}\). Again, \(N_\mathrm{H^{13}CN}/N_\mathrm{H^{13}CO^{+}}=3.3\), yielding \(N_\mathrm{H^{13}CN}=4.6\times10^{13}~\mathrm{cm}^{-2}\), and, assuming a non-enhanced abundance of HCN, \(N_\mathrm{HCN}=2.3\times10^{15}~\mathrm{cm}^{-2}\). Setup 3 adopted \([^{12}\mathrm{C}]/[^{13}\mathrm{C}]=20\), so that the column densities are \(N_\mathrm{H^{13}CN}=1.2\times10^{14}~\mathrm{cm}^{-2}\) and \(N_\mathrm{H^{13}CO^{+}}=3.5\times10^{13}~\mathrm{cm}^{-2}\); other parameters are unchanged.

By contrast, setup 4 adopted an enhanced abundance of HCN relative to HCO\(^{+}\). We kept the same values of \(N_\mathrm{HCO^{+}}=7.0\times10^{14}~\mathrm{cm}^{-2}\) and \(N_\mathrm{H^{13}CN}/N_\mathrm{H^{13}CO^{+}}=3.3\), but assumed that \(N_\mathrm{HCN}/N_\mathrm{HCO^{+}}=6.6\), i.e., that there is an enhancement by a factor of \(\sim2\) (\(N_\mathrm{HCN}=4.6\times10^{15}~\mathrm{cm}^{-2}\)). (Increasing \(N_\mathrm{H^{13}CN}/N_\mathrm{H^{13}CO^{+}}\) too by the same factor resulted in a much less constrained solutions.) High abundance ratios have been proposed to explain the elevated \(R_\mathrm{H}\) in a number of active galactic nuclei (e.g., \citealt{Izu13,Miy17}). Judging from the trends in the results of setups 1-3, we set the ratio \([^{12}\mathrm{C}]/[^{13}\mathrm{C}]=25\), which yielded \(N_\mathrm{H^{13}CO^{+}}=2.8\times10^{13}~\mathrm{cm}^{-2}\) and \(N_\mathrm{H^{13}CN}=9.2\times10^{13}~\mathrm{cm}^{-2}\).

\begin{figure}
\epsscale{0.8}
\plotone{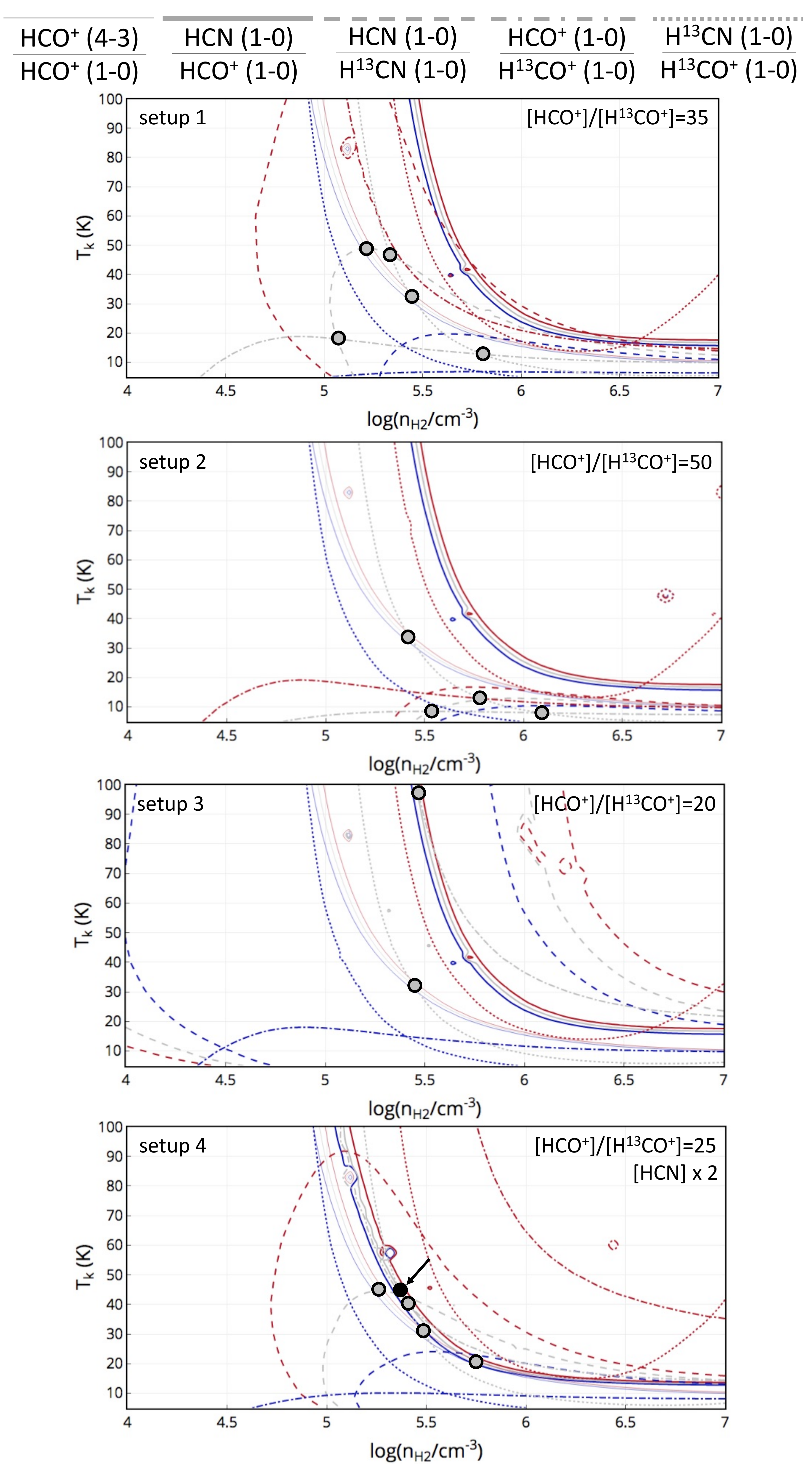}
\caption{Line intensity ratios in the CND derived by RADEX (setups 1-4 in text). The gray curves are the line ratios and the red/blue curves are the upper/lower limits derived from statistical uncertainties. The circles mark the positions where the line intensity ratios intersect; the filled circle indicated by an arrow in setup 4 encloses five intersection points.\label{fig:radex0}}
\end{figure}

The results of the calculations of all four setups are shown in Figure \ref{fig:radex0}. Given that five line intensity ratios are used, there can be ten intersections, and the quality of the chosen setup was evaluated based on the number of intersections and how close they appear in the investigated parameter space. Figure \ref{fig:radex0} shows that, although there are intersections in all investigated setups, setup 4 provides the strongest constraints on physical conditions (the line intensity ratios are best confined in the parameter space). There are nine intersections in setup 4, out of which five coincide in the parameter space, yielding a range of densities of \(10^{4.9}~\mathrm{cm^{-3}}\lesssim n_\mathrm{H_2}\lesssim 10^{5.8}~\mathrm{cm^{-3}}\) and a kinetic temperature range of \(20~\mathrm{K}\lesssim T_\mathrm{k}\lesssim100~\mathrm{K}\) (on average \(T_\mathrm{k}\sim40~\mathrm{K}\)), where the lower and upper limits are dominated by \(r_\mathrm{HCN}\), but also sensitive to \(R_\mathrm{H}\) and \(r_\mathrm{H}\). This result implies the presence of dense and warm molecular gas in the CND. Since \(R_\mathrm{HCO^{+}}\) includes the \(J=4-3\) line, it is possible that the low and high-transition lines of HCO\(^{+}\) arise from somewhat different volumes; selective dissociation may also reduce the emitting volumes of the \(^{13}\)C-bearing molecules. If that is the case, the result may be affected to some extent by a difference in beam filling factors. Note that the three most separated intersections in Figure \ref{fig:radex0} (setup 4) lie on \(R_\mathrm{HCO^{+}}\). Increasing the ratio (e.g., correcting for a low beam filling factor of \(J=4-3\)), generally shifts the ratio toward higher values of \(n_\mathrm{H_2}\) and \(T_\mathrm{k}\); for slightly higher ratios of \(R_\mathrm{HCO^{+}}\sim0.5\), all intersections shift closer together near \(T_\mathrm{k}\sim45\) K and \(n_\mathrm{H_2}\sim10^{5.4}\) cm\(^{-3}\).

\subsection{Calculations using CO (3-2)/CO (1-0)}\label{Db}

\begin{figure}
\epsscale{1.1}
\plotone{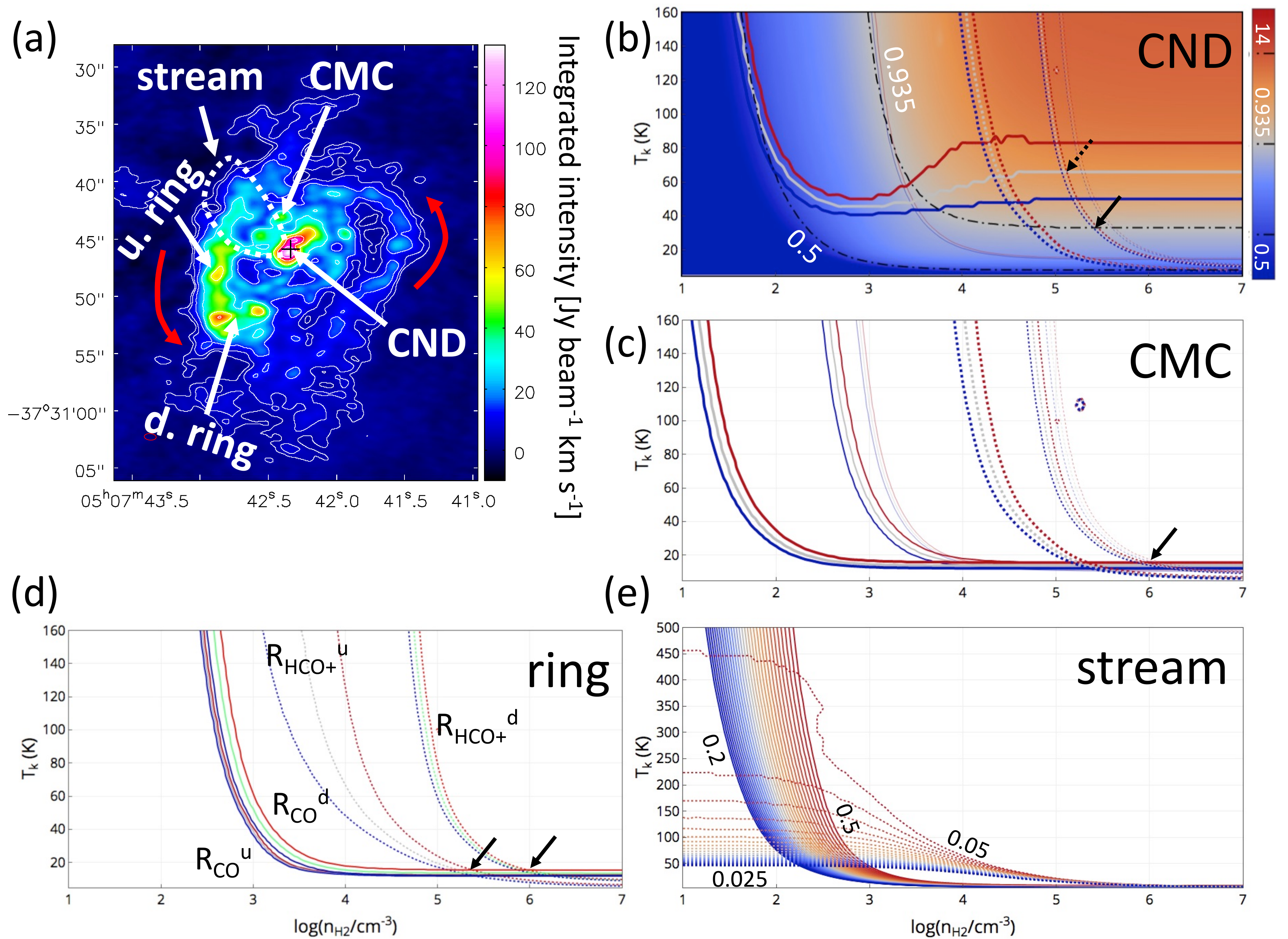}
\caption{Gas conditions derived by RADEX for the regions illustrated on a CO (3-2) integrated intensity map in panel (a). The red arrows indicate galactic rotation and the CO (3-2) contours are plotted at \((0.03,0.05,0.1,0.2,0.4,0.6,0.8)\times133.4~\mathrm{Jy~beam^{-1}~km~s^{-1}}\). The solid and dot-dashed curves are \(R_\mathrm{CO}\) and dotted curves are \(R_\mathrm{HCO^{+}}\); the values of the upper and lower limits are shown. For the CND and CMC (panels b and c), the curves are the results from \(N_\mathrm{CO}=7\times10^{17}~\mathrm{cm}^{-2}\) (thin solid line), \(N_\mathrm{CO}=7\times10^{18}~\mathrm{cm}^{-2}\) (dot-dashed; color), and \(N_\mathrm{CO}=7\times10^{19}~\mathrm{cm}^{-2}\) (thick), and equivalently \(N_\mathrm{HCO^{+}}=7\times10^{13}~\mathrm{cm}^{-2}\) (thin dotted line), \(N_\mathrm{HCO^{+}}=7\times10^{14}~\mathrm{cm}^{-2}\) (normal), and \(N_\mathrm{HCO^{+}}=7\times10^{15}~\mathrm{cm}^{-2}\) (thick) for HCO\(^{+}\). (d) Results for \(N_\mathrm{CO}=7\times10^{18}~\mathrm{cm}^{-2}\) and \(N_\mathrm{HCO^{+}}=7\times10^{14}~\mathrm{cm}^{-2}\) for the upstream ``u'' and downstream ``d'' sides of the ring. (e) Results for the stream region (nuclear outflow). The black arrows in panels (b)-(e) indicate the intersection areas obtained for moderate column densities. The dashed arrow in panel (a) marks a hybrid point that corresponds to \([\mathrm{HCO^{+}}]/[\mathrm{CO}]=10^{-5}\).\label{fig:radex}}
\end{figure}

The analysis is now extended to include the line intensity ratios of CO, which are available throughout the central 1 kpc region (Figure \ref{fig:radex}(a)). The calculations for the CND were carried out for different parameters as follows (Figure \ref{fig:radex}(b)): CO column density ranging from \(N_\mathrm{CO}=7\times10^{17}~\mathrm{cm}^{-2}\) to \(N_\mathrm{CO}=7\times10^{19}~\mathrm{cm}^{-2}\) and HCO\(^{+}\) column density ranging from \(N_\mathrm{HCO^+}=7\times10^{13}~\mathrm{cm}^{-2}\) to \(N_\mathrm{HCO^+}=7\times10^{15}~\mathrm{cm}^{-2}\) (assuming an abundance ratio of \([\mathrm{HCO^{+}}]/[\mathrm{CO}]=10^{-4}\)). A constant velocity width of \(\Delta V=50~\mathrm{km~s}^{-1}\) was adopted. The results, summarized in the first row of Table \ref{tab4}, where \(T_\mathrm{k}=40\) K was assumed, are generally consistent with Figure \ref{fig:radex0}, indicating gas at high pressure of \(10^4\mathrm{-}10^7~\mathrm{cm^{-3}~K}\), higher than typical gas pressure in Galactic disk clouds and comparable to the environment in the Galactic center region. Figure \ref{fig:radex}(a) (dashed arrow) shows that for a larger column density of CO, \(N_\mathrm{CO}=7\times10^{19}\) cm\(^{-2}\) and \([\mathrm{HCO^{+}}]/[\mathrm{CO}]=10^{-5}\), we obtain higher temperatures (\(50\mathrm{-}90\) K), also consistent with the derived values in section \ref{Daa}. The distribution of \(R_\mathrm{CO}\), indicated by color in panel (b), also shows that, in general, higher values of \(R_\mathrm{CO}\) result in higher densities and/or temperatures. If the beam filling factor of CO is larger than that of HCO\(^{+}\), correcting for this effect would enhance \(R_\mathrm{CO}\) and shift the solutions toward the regime of higher densities and/or temperatures, consistent with results in the previous section.

\begin{figure}
\epsscale{1.1}
\plotone{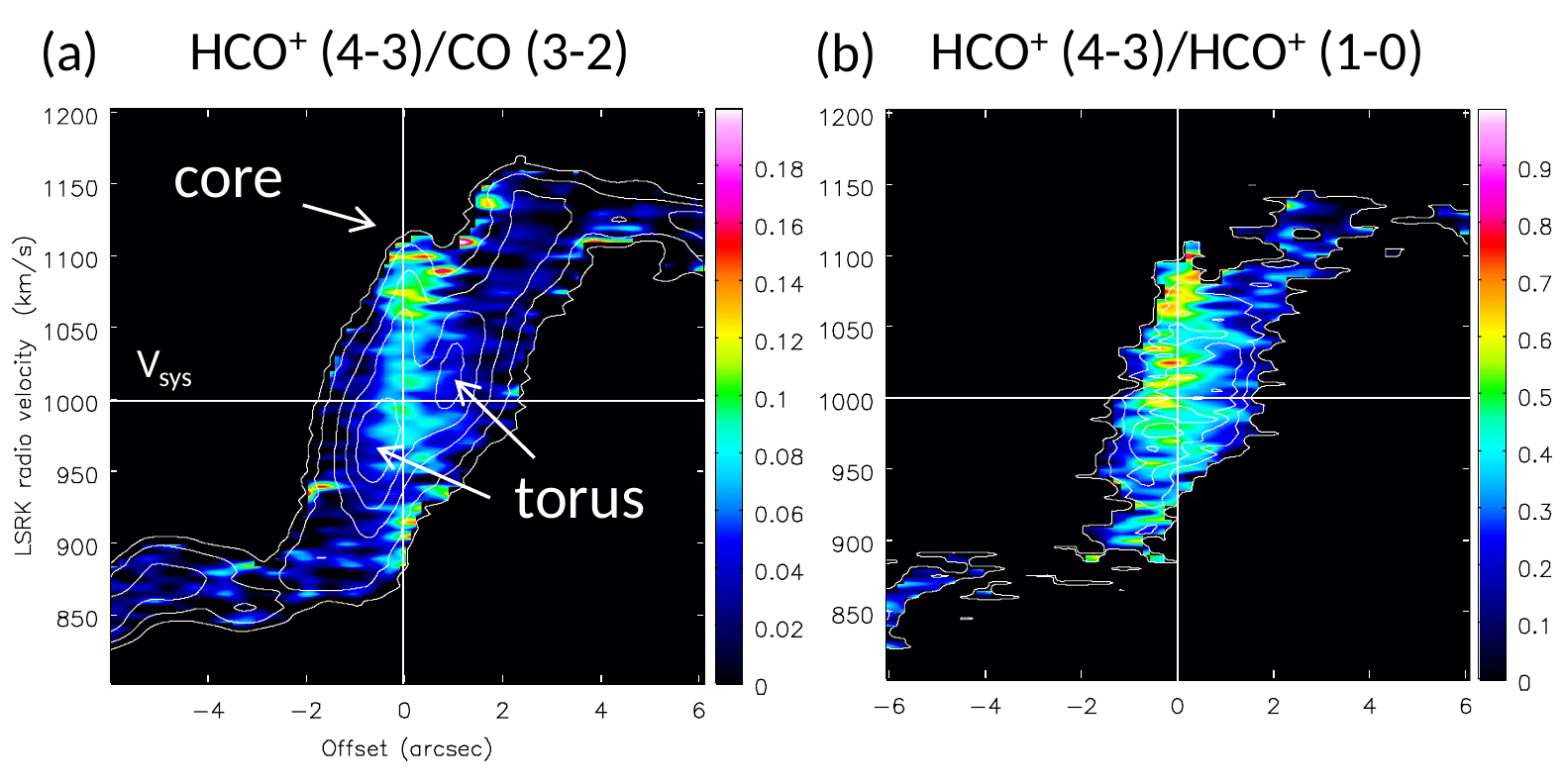}
\caption{(a) HCO\(^{+}\) (4-3)/CO (3-2) intensity ratio in position-velocity space with CO (3-2) contours. (b) HCO\(^{+}\) (4-3)/HCO\(^{+}\) (1-0) intensity ratio with HCO\(^{+}\) (1-0) contours. The data were adjusted to a comparable resolution prior to calculating the ratios.\label{fig:pvdr}}
\end{figure}

Note that the highest measured excitation of HCO\(^{+}\) molecules, reflected in the elevated ratio \(R_\mathrm{HCO^{+}}\sim0.6\), is found toward the core inside the CND (the high-velocity component shown in Figure \ref{fig:pvdr}(b)). The core also exhibits an elevated ratio of HCO\(^{+}\) (4-3) to CO (3-2) compared to the rest of the CND (panel (a)) and it is spatially coincident with the 93 GHz continuum (Figure \ref{fig:linrat}(a)). The continuum has been carefully subtracted so that it does not affect the HCO\(^{+}\) (4-3) spectrum significantly; this is also supported by the absence of emission in the HCO\(^{+}\) (4-3) spectrum at velocities \(V_\mathrm{LSR}<900\) km s\(^{-1}\) and \(V_\mathrm{LSR}>1120\) km s\(^{-1}\) in the PVD diagram in Figure \ref{fig:torus-dense} (middle panel in the bottom row). The same plot also shows enhanced HCO\(^{+}\) (4-3) emission in the ``core'' component relative to the surrounding gas torus. The behavior is clearly different from that of HCO\(^{+}\) (1-0) shown in the second row. For the ratios of \(R_\mathrm{CO}=1.50\pm0.50\) and \(R_\mathrm{HCO^{+}}=0.50\pm0.12\) and the same column densities as adopted above for the CND, RADEX calculations yield a temperature range of 40-70 K and gas densities of \(\log{(n_\mathrm{H_2}/\mathrm{cm^{-3}})}=5.0\mathrm{-}5.6\), within uncertainties. Here, we have assumed that \(1<R_\mathrm{CO}<2\) and that \(R_\mathrm{CO}\) and \(R_\mathrm{HCO^{+}}\) trace the same volumes.

The CMC consists of regions R1 and R2, which represent giant molecular clouds where all gas tracers are firmly detected (Figure \ref{fig:radex}). RADEX calculations based on \(R_\mathrm{HCO^{+}}\) yield gas densities comparable to those in the CND for an assumed temperature of 30 K. On the other hand, a derivation based on \(R_\mathrm{CO}\) for the same temperature results in two orders of magnitude lower densities (Table \ref{tab4}). \(R_\mathrm{CO}\) and \(R_\mathrm{HCO^{+}}\) cross at a point that corresponds to a relatively high density of \(\sim10^6\) cm\(^{-3}\) and a low temperature of \(\sim15\) K. Considering the results for the CND in section \ref{Daa}, we note that the intersection of \(R_\mathrm{CO}\) and \(R_\mathrm{HCO^{+}}\) may be an artificial solution produced by a difference in beam filling factors (emitting volumes). If that is the case, the excitation of the dense gas tracers HCN and HCO\(^{+}\) reflects the inner cloud regions, whereas CO, whose optical depth is larger and critical density lower, traces the conditions in the outer regions and envelopes.

The investigated part of the 500 pc ring consists of regions R3 and regions R4 and R5. These are further classified based on their location with respect to the ring rotation about the galactic center (see panel (a) in Figure \ref{fig:radex}): R3 is located ``upstream'' whereas R4 and R5 (whose average is used in calculations) are ``downstream.'' The results in Figure \ref{fig:radex} and Table \ref{tab4} suggest that either gas density or temperature rises inside the ring from the upstream to the downstream side; this can be explained as a result of enhanced star formation activity within the ring, as discussed in \cite{Sal17} in terms of cloud evolution. The clouds enter the 500 pc ring from the large-scale bar (upstream side) and engage in starburst activity as they revolve around the galactic center. The downstream regions R4 and R5 are located at the point of highest star forming activity in the ring, as revealed by 93 GHz continuum tracing free-free emission.

\subsection{Stream (nuclear outflow)}\label{Dc}

\begin{figure}
\epsscale{1}
\plotone{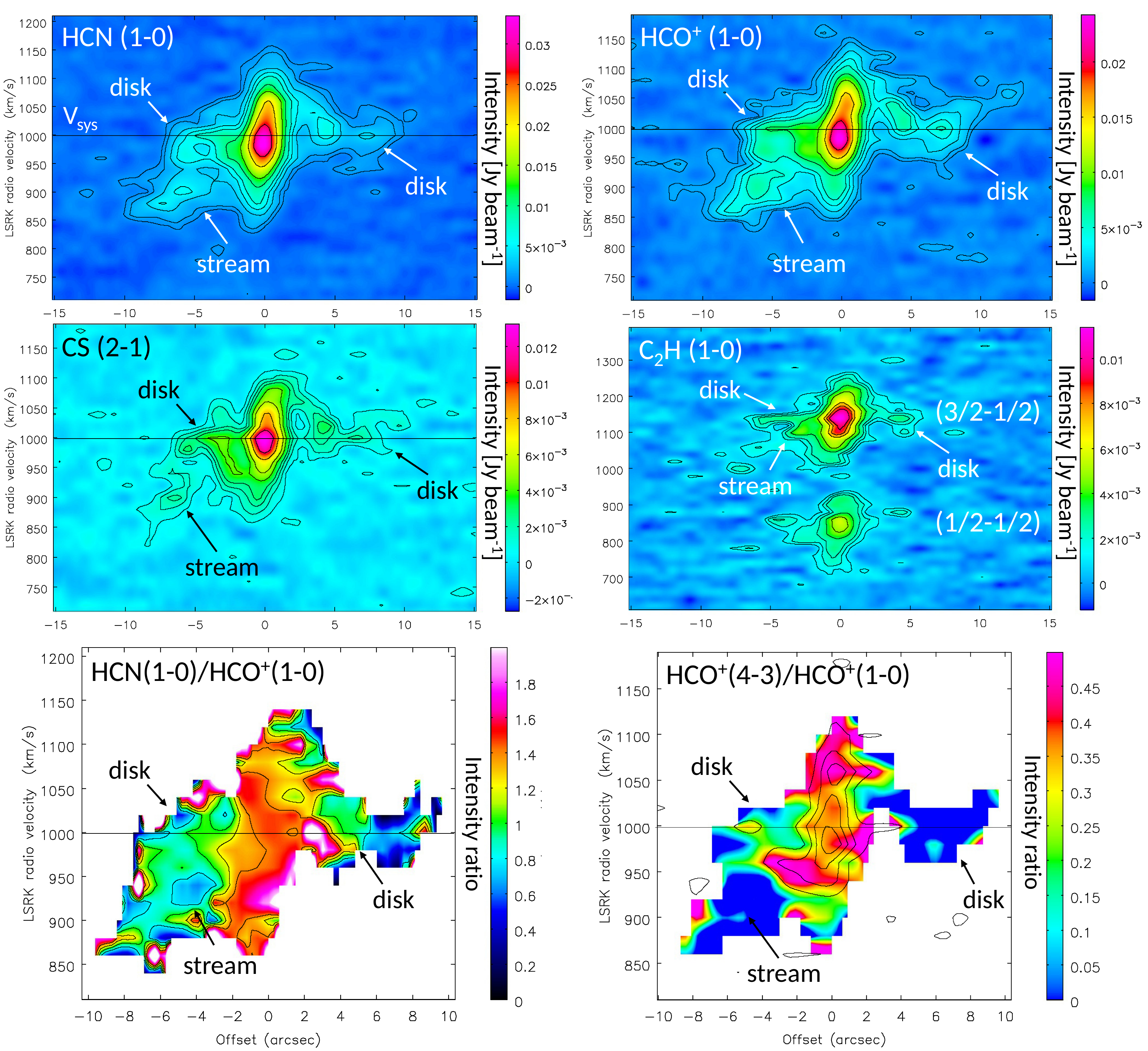}
\caption{Position-velocity diagrams (PVDs) of HCN (1-0), HCO\(^{+}\) (1-0), CS (2-1), and C\(_2\)H (1-0) along the minor galactic axis at \(\mathrm{PA}=54\arcdeg\). High-sensitivity data cubes (natural weighting) were used to derive the PVDs. The intensity contours are plotted at \((0.03, 0.05, 0.1, 0.15, 0.2, 0.3, 0.4, 0.6, 0.7, 0.8)\times\mathcal{S}_\mathrm{max}\) where \(\mathcal{S}_\mathrm{max}^\mathrm{HCN(1-0)}=33.5~\mathrm{mJy~beam^{-1}}\), \(\mathcal{S}_\mathrm{max}^\mathrm{HCO^{+}(1-0)}=24.3~\mathrm{mJy~beam^{-1}}\), \(\mathcal{S}_\mathrm{max}^\mathrm{CS(2-1)}=13.3~\mathrm{mJy~beam^{-1}}\), and \(\mathcal{S}_\mathrm{max}^\mathrm{C_2H(1-0)}=11.4~\mathrm{mJy~beam^{-1}}\). The lowest contour of the C\(_2\)H (1-0) image is 0.1 \textbf{(\(3~\sigma\))}, and the LSR velocity is not corrected for the fine structure components; the rest frequency corresponds to the frequency between the two multiplets.. Both triplets (\(J=3/2-1/2\) and \(J=1/2-1/2\)) are plotted. The bottom left panel shows the HCN (1-0) to HCO\(^{+}\) (1-0) line intensity ratio \(R_\mathrm{H}\) with contours plotted at \(0.8,1.0,1.2,1.4\). The bottom right panel shows the HCO\(^{+}\) intensity ratio PVD with HCO\(^{+}\) (4-3) contours at \((0.2,0.4,0.6,0.8)\times0.175\) Jy beam\(^{-1}\).\label{fig:pvdoutdense}}
\end{figure}

To estimate the physical conditions in the nuclear outflow (labeled ``stream'' in Figure \ref{fig:radex}(a)), we derived the ratios of \(R_\mathrm{CO}\) and \(R_\mathrm{HCO^{+}}\) in position-velocity space along the minor galactic axis, which is assumed to be the direction of the outflow \citep{Sal16}. The \(R_\mathrm{CO}\) ratio was presented in \cite{Sal17}, while in Figure \ref{fig:pvdoutdense} (bottom right panel) we present the \(R_\mathrm{HCO^{+}}\) ratio derived at a comparable angular resolution of \(\sim2\arcsec\). For low column densities of \(N_\mathrm{CO}=1\times10^{18}~\mathrm{cm}^{-2}\) (equivalent to optically thin CO (1-0) emission; see section \ref{Da}) and \(N_\mathrm{HCO^{+}}=1\times10^{14}~\mathrm{cm}^{-2}\), assumed temperature of \(T_\mathrm{k}>30\) K, and a velocity width of \(\Delta V=50~\mathrm{km~s}^{-1}\), corresponding to the velocity with of the extraplanar stream, RADEX calculations yielded relatively low densities as listed in Table \ref{tab4} and plotted in Figure \ref{fig:radex}(e). However, since HCO\(^{+}\) (4-3) is not clearly detected in the outflow, the ratio \(R_\mathrm{HCO^{+}}\) is not well constrained; we plot the results in the range \(0.025<R_\mathrm{HCO^{+}}<0.05\). The resulting parameters are notably different from those in other investigated regions in the starburst disk. In particular, the density, which is better constrained by \(R_\mathrm{CO}\) than temperature, is lower than in the massive molecular clouds in the starburst disk. Since the density was derived over 100-pc regions, smaller structures are not resolved. If the physical conditions of the outflow gas are uniform, then the result reflects diffuse (\(n_\mathrm{H_2}\sim10^2-10^3~\mathrm{cm}^{-3}\)) and warm (\(T_\mathrm{k}\gtrsim20~\mathrm{K}\)) medium, comparable to the conditions in the super-bubble in M82 \citep{CM16}. On the other hand, the same result can also imply a clumpy gas, where the clumps of dense gas are unresolved within the probed aperture. The clumpy nature is likely given the detection of other dense gas tracers in the stream, namely, HCN, CS, and C\(_2\)H presented below. All these tracers have critical densities of the order \(n_\mathrm{cr}\gtrsim10^3~\mathrm{cm}^{-3}\) in the regime of the most likely kinetic temperatures. \cite{Sal17} also showed that \(R_\mathrm{CO}\) is decreasing with galactocentric distance in extraplanar gas, from 0.5 near the CND to 0.2 at the farthermost point. Figure \ref{fig:radex}(e) implies that, for a constant column density, such trend reflects a decrease in gas density (divergence of mass flux or molecule destruction) and/or temperature in the outflow direction.

Also shown in Figure \ref{fig:pvdoutdense} are the PVDs along the minor galactic axis for HCN (1-0), HCO\(^{+}\) (1-0), CS (2-1), and C\(_2\)H (1-0). Dense gas tracers, such as these, were recently detected in the molecular outflow in the starburst galaxy NGC 253 \citep{Wal17}, providing evidence that starburst-driven outflows can include dense gas. The line intensity ratio of HCN (1-0) to HCO\(^{+}\) (1-0), \(R_\mathrm{H}\), in position-velocity space is shown on the bottom left panel of the figure. The PVDs of these lines show evidence of dense molecular gas in the outflow with kinematics similar to that of CO. In particular, the PVDs exhibit an outward velocity gradient reported earlier by \cite{Sal17} as a possible evidence of outflow acceleration. Note that \(R_\mathrm{H}>1\) in the CND, and decreases to \(\sim1\) at a radius \(r\gtrsim5\arcsec\) (260 pc). There is signature of local enhancements (\(R_\mathrm{H}>1\)) in some regions of the outflow, especially at the high velocity end between \(\mathrm{offset}=-4\arcsec\) and \(0\arcsec\), and between \(V_\mathrm{LSR}=850~\mathrm{km~s}^{-1}\) and \(900~\mathrm{km~s}^{-1}\), but the ratio is close to unity (the mean ratio in a region defined by \(\mathrm{offset}<-2\arcsec\) and \(V_\mathrm{LSR}<950~\mathrm{km~s}^{-1}\) is \(R_\mathrm{H}^\mathrm{stream}=1.1\pm0.8\)), implying that the HCN (1-0) to HCO\(^{+}\) (1-0) ratio in the outflow is similar to the average value in the starburst disk and lower than the ratio in the CND. If the stream gas is ejected from the CND, the decrease of \(R_\mathrm{H}\) along the outflow direction can imply that the electron densities in the outflow are significantly lower than in the CND, yielding a decreased effect of electron excitation and hence smaller \(R_\mathrm{H}\).

\begin{figure}
\epsscale{0.5}
\plotone{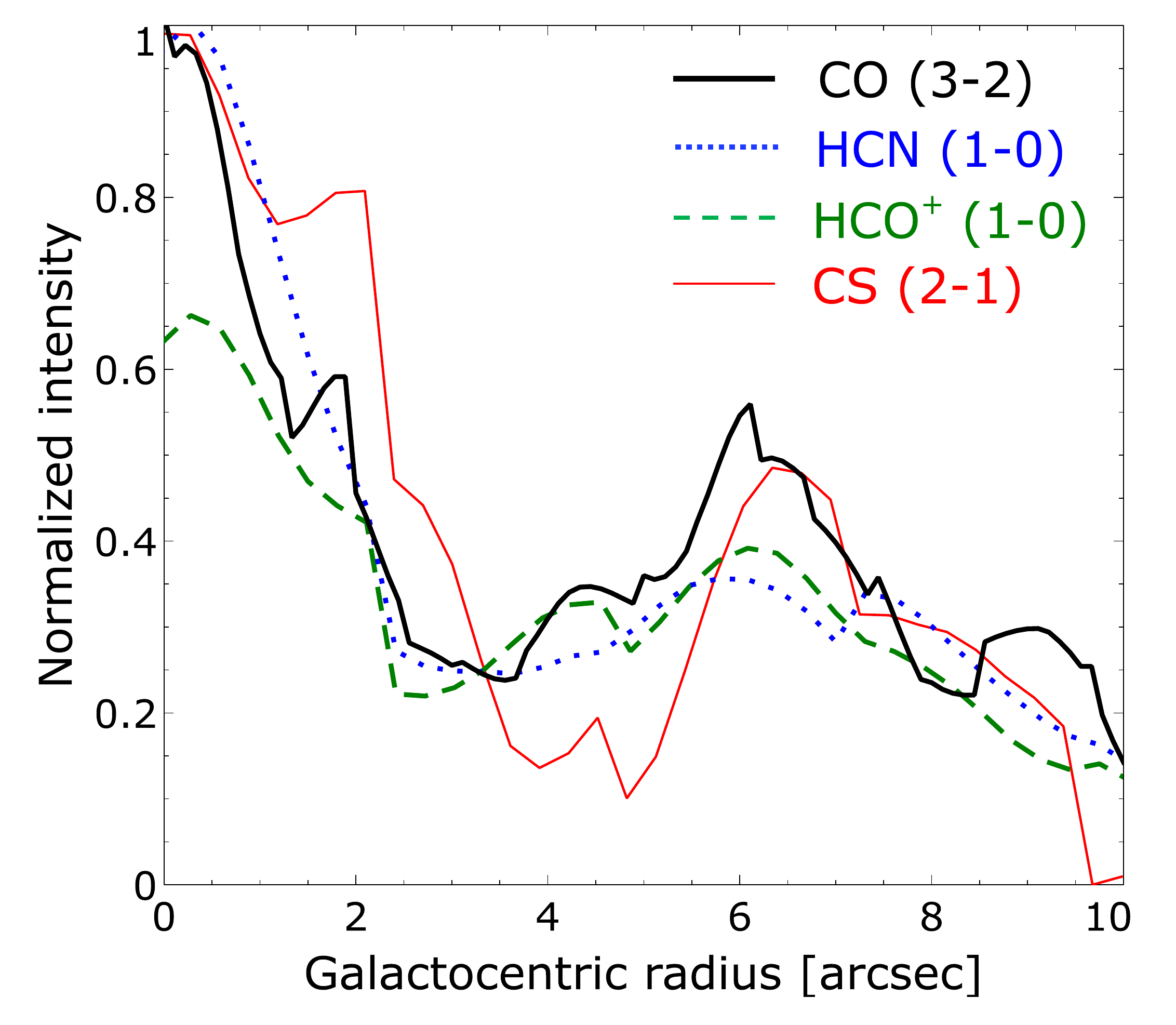}
\caption{Normalized intensities of the dense gas tracers CO (3-2), HCN (1-0), HCO\(^{+}\) (1-0), and CS (2-1) along the stream defined from \((Y,V_\mathrm{LSR})=(0\arcsec,905~\mathrm{km~s}^{-1})\) to \((Y,V_\mathrm{LSR})=(-10\arcsec,865~\mathrm{km~s}^{-1})\), where \(Y\) is parallel to the minor galactic axis. The intensities of CO (3-2) and CS (2-1) are normalized to that of HCN (1-0). The rms noise levels are \(4\%\) (CO), \(6\%\) (HCN), \(7\%\) (HCO\(^{+}\)), and \(20\%\) (CS) of the maximum intensity.\label{fig:ofl}}
\end{figure}

The middle rows of Figure \ref{fig:pvdoutdense} show that the emission of CS (2-1) and C\(_2\)H (1-0) arises primarily from the clouds in the galactic plane rotating at velocities close to the systemic velocity \(V_\mathrm{sys}\), denoted by ``disk''. However, the line splitting into ``disk'' and ``stream'' (outflow) in the figure, indicates the presence of CS and C\(_2\)H in the extraplanar gas. While CS (2-1) is detected at \(4~\sigma\) at \((\mathrm{offset},V_\mathrm{LSR})=(-7'',900~\mathrm{km~s}^{-1})\), the emission of C\(_2\)H (1-0) in the stream component is detected at \(3~\sigma\). Figure \ref{fig:ofl} shows that the intensity projections of the dense gas tracers in the outflow direction is similar to that of CO (3-2); the major differences are observed in the HCN (1-0) to HCO\(^{+}\) (1-0) ratio, which decreases from \(\sim1.4\) in the CND to \(\sim1\) in the stream, and the intensity of CS (2-1) which is enhanced at the radius of \(r\sim2\arcsec\) and relatively low at \(r\sim5\arcsec\), though within the uncertainty of \(20\%\).

\begin{table}
\begin{center}
\caption{Physical Conditions Derived with RADEX}\label{tab4}
\begin{tabular}{lcccccc}
\tableline\tableline
Region & \(N_\mathrm{CO}\) (cm\(^{-2}\)) & \(T_\mathrm{k}~(\mathrm{K})\) & \(\log{(n_\mathrm{H_2}/\mathrm{cm}^{-3})}\) & \(\log{(n_\mathrm{H_2}/\mathrm{cm}^{-3})}\) \\
& & & \((R_\mathrm{CO})\) & \((R_\mathrm{HCO^{+}})\) & & \\
\hline
CND & \(7\times10^{18}\) & \(40\) & 3.7-4.0 & 5.3-5.4 \\
CMC & \(4\times10^{18}\) & 30 & 3.2-3.5 & 5.4-5.6  \\
500 pc ring (upstream) & \(5\times10^{18}\) & 30 & 3.0-3.2 & 4.5-4.8 \\
500 pc ring (downstream) & \(4\times10^{18}\) & 30 & 3.2-3.5 & 5.4-5.6 \\
Stream (nuclear outflow) & \(1\times10^{18}\) & \(>30\) & 1.3-3.0 & ...  \\
\tableline
\end{tabular}
\end{center}
\end{table}

\subsection{Uncertainties of the radiative transfer analysis}\label{Dd}

There are several causes of systematic errors that dominate the overall uncertainty of the non-LTE analysis presented above. First, the emission from the investigated regions is regarded as an average within an aperture size of \(2\arcsec\) (100 pc). Since this is probably larger than the size of individual giant molecular clouds (e.g., \citealt{SSS85}), the line intensity reflects an average over a cloud/envelope/inter-cloud region. The physical conditions and chemical composition may vary from the cloud to the inter-cloud medium.

Second, in one part of the analysis (section \ref{Db}), some solutions were based on an assumption that CO and HCO\(^{+}\) emission arises from the same cloud volumes. However, as mentioned above, even the lines of the same species may be tracing somewhat different regions. For example, while HCO\(^{+}\) (4-3) may be tracing cloud interior, HCO\(^{+}\) (1-0) can be more prominent in the outer layers (e.g., \citealt{Hog97}). In other words, the beam filling factor may vary among the tracers. In general, the solutions of physical conditions based on mixed \(R_\mathrm{CO}\) and \(R_\mathrm{HCO^{+}}\) ratios produce lower kinetic temperature (10-20 K). When the ratios are considered separately, and \(^{13}\)C-bearing molecular species are included, the dense gas is found to be warmer (CND; section \ref{Daa}). Considering that the CND and CMCs are dominated by vigorous star formation and its feedback (e.g., shock heating), higher temperatures are plausible.

Also, the column densities of CO and HCO\(^{+}\) were estimated from indirect methods (CO luminosity converted to H\(_2\) column density by using a constant conversion factor, and then converted to CO and HCO\(^{+}\) column densities by assuming abundance ratios). For this purpose, we estimated the CO column density in the case of optically thin CO (1-0) emission, to obtain a lower limit. The column density of HCO\(^{+}\) was also derived by assuming LTE. The uncertainty of the column density is estimated to be a factor of a few.

The results obtained in the analysis are generally consistent with previous studies of gas conditions in NGC 1808 based on single pointing data from single dish telescopes \citep{Aal94,Sal14}. Measuring the intensity ratios from additional CO lines, including optically thin lines of \(^{13}\)CO and C\(^{18}\)O, as well as other dense gas tracers with comparable critical densities throughout the starburst disk (compact and diffuse components) would be a next step of the analysis to obtain a more detailed picture of the physical conditions of molecular medium.

\section{Summary}\label{E}

We have presented multi-line observations of dense molecular gas in the central 1 kpc of the nearby starburst galaxy NGC 1808 carried out with ALMA. The main findings and conclusions are summarized below.

\begin{enumerate}

\item{The following spectral lines were detected for the first time in NGC 1808 in ALMA cycle 2 observations: H\(^{13}\)CN (1-0), H\(^{13}\)CO\(^{+}\) (1-0), SiO (2-1), C\(_2\)H (1-0), HOC\(^{+}\) (1-0), CS (2-1), and HCO\(^{+}\) (4-3). In addition, HCN (1-0) and HCO\(^{+}\) (1-0) are firmly detected throughout the central 1 kpc region at unprecedented angular resolution of \(\sim50\) pc. The intensity distributions of dense gas tracers in the CND (central 100 pc) are consistent with a torus (double peak) structure previously revealed with CO (3-2) data, and an unresolved core. We also report tentative detections of SO (2-1), HN\(^{13}\)C (1-0), and HNCO (4-3).}

\item{The line intensities of HCN (1-0) and C\(_2\)H (1-0) are relatively enhanced with respect to HCO\(^{+}\) (1-0), CS (2-1), and CO (1-0) in the CND region. The intensity ratio of HCN (1-0) to HCO\(^{+}\) (1-0) in the CND is observed to be \(R_\mathrm{H}\equiv W_\mathrm{HCN(1-0)}/W_\mathrm{HCO^{+}(1-0)}\sim1.5\), consistent with observations of composite starburst plus AGN nuclei in nearby galaxies. There is evidence of an enhanced ratio in a star-forming ring (including molecular spiral arms) at a radius of \(\sim300\) pc around the CND. The enhanced ratio \(R_\mathrm{H}\) shows tentative spatial correlation with SiO (2-1) and radio continuum tracing shocks, free-free emission from H II regions, and synchrotron emission from supernova remnants. The results indicate that the ratio is affected by star formation activity, possibly in terms of shock heating and electron excitation, which can be tested with new observations at higher resolution. The producers of shocks and electrons in the starburst disk are supernova explosions and H II regions.}

\item{The H\(_2\) gas densities and kinetic temperatures (averaged over 100 pc) in the massive molecular clouds of the CND, CMCs, and 500 pc ring were estimated under non-LTE analysis using the radiative transfer program RADEX. The calculations typically yield dense (\(n_\mathrm{H_2}\sim10^5\) cm\(^{-3}\)) and warm (\(20~\mathrm{K}\lesssim T_\mathrm{k}\lesssim100~\mathrm{K}\)) molecular gas in the CND. The CMC and 500 pc ring regions exhibit lower gas excitation, with evidence of an increase in temperature or density from the upstream to the downstream side in the ring. The average density in the outflow (\(\sim10^2-10^3\) cm\(^{-3}\)) is an order of magnitude lower than in the massive clouds in the starburst disk. The highest excitation of HCO\(^{+}\) gas, expressed as a HCO\(^{+}\) (4-3)/HCO\(^{+}\) (1-0) line intensity ratio of \(\sim0.6\) is found in an unresolved core that exhibits peculiar kinematics within the CND.}

\item{The lines of HCN (1-0), HCO\(^{+}\) (1-0), CS (2-1), and C\(_2\)H (1-0) were detected for the first time in the superwind of NGC 1808. All dense gas tracers exhibit kinematics similar to that of CO (3-2) reported earlier, supporting the presence of a velocity gradient in the direction of the outflow. Their line intensity ratios are generally similar to those in the central 1 kpc starburst disk.}

\end{enumerate}

\acknowledgements

The authors humbly thank the anonymous referee for insightful comments and suggestions. This paper makes use of the following ALMA data: ADS/JAO.ALMA\#2012.1.01004.S, ADS/JAO.ALMA\#2013.1.00911.S. ALMA is a partnership of ESO (representing its member states), NSF (USA), and NINS (Japan), together with NRC (Canada) and NSC and ASIAA (Taiwan), in cooperation with the Republic of Chile. The Joint ALMA Observatory is operated by ESO, AUI/NRAO, and NAOJ. The National Radio Astronomy Observatory is a facility of the National Science Foundation operated under cooperative agreement by Associated Universities, Inc. Based on observations made with the NASA/ESA \emph{Hubble Space Telescope} and obtained from the Hubble Legacy Archive, which is a collaboration between the Space Telescope Science Institute (STScI/NASA), the Space Telescope European Coordinating Facility (ST-ECF/ESA), and the Canadian Astronomy Data Centre (CADC/NRC/CSA). This research has made use of the NASA/IPAC Extragalactic Database (NED), which is operated by the Jet Propulsion Laboratory, California Institute of Technology, under contract with the National Aeronautics and Space Administration. D.S. was supported by the ALMA Japan Research Grant of NAOJ Chile Observatory, NAOJ-ALMA-181.

\appendix

\section{Optical depth and column density}\label{AP}

Assuming local thermodynamic equilibrium (LTE), i.e., conditions where all transitions between energy states in molecules are determined by a single excitation temperature \(T_\mathrm{ex}\), it is possible to calculate the optical depth of the dense gas tracers for which the lines of two or more isotopes are available. The brightness temperature of a source, \(T_\mathrm{b}\), can be expressed as

\begin{equation}
T_\mathrm{b}=\left[J(T_\mathrm{ex})-J(T_\mathrm{bg})\right]\left(1-e^{-\tau}\right),
\end{equation}
where \(T_\mathrm{ex}\) is the excitation temperature, \(T_\mathrm{bg}\) is the background temperature, \(\tau\) is the optical depth of the spectral line, and

\begin{equation}
J(T)=\frac{h\nu}{k}\frac{1}{e^{h\nu/kT}-1}
\end{equation}
is the radiation temperature, where \(h\) is the Planck constant, and \(k\) is the Boltzmann constant. The ratio of the brightness temperatures of two lines (assuming similar excitation temperatures and line frequencies) is simply

\begin{equation}\label{opt}
\frac{T_\mathrm{b,1}}{T_\mathrm{b,2}}=\frac{1-e^{-\tau_1}}{1-e^{-\tau_2}}=\frac{1-e^{-\tau_2/\mathcal{R}}}{1-e^{-\tau_2}},
\end{equation}
where \(\mathcal{R}\equiv\tau_2/\tau_1\).

The total column density of (rigid) linear molecules, \(N\), can be calculated under LTE approximation (see, e.g., Appendix in \citealt{Gar91}) from

\begin{equation}\label{column}
N=\frac{3k}{8\pi^3 B \mu^2}\frac{\exp(E_J/kT_\mathrm{ex})}{J+1}\frac{T_\mathrm{ex}+hB/3k}{1-\exp(-h\nu/kT_\mathrm{ex})}\int\tau dv
\end{equation}
where \(J\) is the rotational quantum number of the lower state in a \(J+1\rightarrow J\) transition, \(E_J=hBJ(J+1)\) is its energy, \(B\) is the rotational constant, \(\mu\) is the permanent electric dipole moment, and integration is done over velocity \(v\) within the spectral line. We simplify the integral in the above equation as \(\int\tau dv\approx\tau\Delta V\), and calculate \(N\) from the derived \(\tau\) and measured \(\Delta V\). Note, however, that there is dependence on excitation temperature \(T_\mathrm{ex}\). In the calculations for HCN (1-0) and HCO\(^{+}\) (1-0) lines below, we use \(T_\mathrm{ex}=31\) K, estimated for \(^{12}\)CO (1-0) \citep{Sal14}, and \(T_\mathrm{ex}=10\) K.

\subsection{HCO\(^{+}\) (1-0)}\label{APa}

In the case of the HCO\(^{+}\) molecule, Equation \ref{opt} can be expressed in terms of intensities as

\begin{equation}\label{opthco+}
\frac{\mathcal{S}_\mathrm{13}}{\mathcal{S}_\mathrm{12}}=\frac{1-e^{-\tau_{12}/\mathcal{R}}}{1-e^{-\tau_{12}}},
\end{equation}
where ``12'' and ``13'' denote the isotopologues H\(^{12}\)CO\(^{+}\) and H\(^{13}\)CO\(^{+}\), respectively. In the above equation, we have taken that \(\nu_{12}\approx\nu_{13}\) and \(T_\mathrm{ex,12}\approx T_\mathrm{ex,13}\). Note that we also consider that the beam filling factors of the two lines are identical. The optical depth ratio is approximately the abundance ratio \(\mathcal{R}_\mathrm{HCO^{+}}\approx[\mathrm{HCO^{+}}]/[\mathrm{H^{13}CO^{+}}]\); we adopt the values of 50, following \cite{San12}, and 25, estimated in section \ref{Daa}. The intensities of the two lines, derived by Gaussian fitting of the line profiles within a circle of diameter \(2\arcsec\) centered at the galactic center, are \(\mathcal{S}_{12}=21.6\pm0.9~\mathrm{Jy~beam^{-1}}\) and \(\mathcal{S}_{13}=1.2\pm0.5~\mathrm{Jy~beam^{-1}}\). Solving Equation \ref{opthco+} numerically yields the optical depths of \(\tau_{12}=2.65\) and \(\tau_{13}=0.05\) for \(\mathcal{R}=50\) and \(\tau_{12}=0.73\) and \(\tau_{13}=0.03\) for \(\mathcal{R}=25\): the HCO\(^{+}\) (1-0) line is moderately optically thick, whereas H\(^{13}\)CO\(^{+}\) (1-0) can be regarded as optically thin.

The total column density of HCO\(^{+}\) molecules is calculated from Equation \ref{column} as

\begin{equation}\label{cd}
N_\mathrm{HCO^{+}}=\frac{3k}{8\pi^3 B \mu^2}\frac{T_\mathrm{ex}+hB/3k}{1-\exp(-h\nu/kT_\mathrm{ex})}\tau \Delta V.
\end{equation}
where we have used \(E_J=0\) (ground state). For the adopted values of \(B=44.5944\) GHz, \(\mu=3.89\) D, \(T_\mathrm{ex}=31\) K, \(\tau=2.65\), and \(\Delta V=136\pm4\) km s\(^{-1}\), where \(B\) and \(\mu\) are taken from the Splatalogue database, we find \(N_\mathrm{HCO^{+}}=(2.19\pm0.06)\times10^{16}\) cm\(^{-2}\), where the uncertainty is estimated only from the uncertainty of the velocity width. A lower excitation temperature of \(T_\mathrm{ex}=10\) K, e.g., yields one order of magnitude smaller column density (Table \ref{ap}). On the other hand, applying \(\Delta V=50\) km s\(^{-1}\) and using the derived lower optical depth would result in one order of magnitude lower \(N_\mathrm{HCO^{+}}\), in agreement with the value applied in section \ref{Daa}, which was based on \([\mathrm{HCO^{+}}]/[\mathrm{H_2}]=10^{-8}\).

\subsection{HCN (1-0)}\label{APb}

The intensities of the HCN (1-0) and H\(^{13}\)CN (1-0) lines are \(\mathcal{S}_{12}=31.0\pm0.9~\mathrm{Jy~beam^{-1}}\) and \(\mathcal{S}_{13}=2.3\pm0.3~\mathrm{Jy~beam^{-1}}\), respectively, and the corresponding optical depths are \(\tau_{12}=3.76\) and \(\tau_{13}=0.07\) for the abundance ratio of \(\mathcal{R}_\mathrm{HCN}\approx[\mathrm{HCN}]/[\mathrm{H^{13}CN}]=50\) and \(\tau_{12}=1.47\) and \(\tau_{13}=0.06\) for the abundance ratio of \(\mathcal{R}_\mathrm{HCN}\approx[\mathrm{HCN}]/[\mathrm{H^{13}CN}]=25\). The same abundance ratio is applied for HCN and HCO\(^{+}\), because it is largely determined by the abundance ratio of \([^{12}\mathrm{C}]/[^{13}\mathrm{C}]\). The HCN (1-0) line too is moderately optically thick toward the galactic center. Since this is the only region where H\(^{13}\)CN (1-0) is detected, the calculation using isotopes is limited to the CND.

Finally, the column density is calculated from Equation \ref{cd}. Inserting  \(B=44.31597\) GHz, \(\mu=2.984\) D (from Splatalogue), \(T_\mathrm{ex}=31\) K, \(\tau=3.76\), and \(\Delta V=137\pm4\) km s\(^{-1}\), the column density of HCN molecules becomes \(N_\mathrm{HCN}=(5.39\pm0.15)\times10^{16}\) cm\(^{-2}\).

The result yields a column density ratio (equivalent to the abundance ratio) of \(N_\mathrm{HCN}/N_\mathrm{HCO^{+}}=2.46\pm0.10\) averaged over the central 100 pc. The derived parameters are summarized in Table \ref{ap}. Note that \(\mu^2\) is about two times larger for HCO\(^{+}\) than for HCN. This means that, under the conditions of similar excitation temperatures, velocity widths, and column densities of HCN (1-0) and HCO\(^{+}\) (1-0), the opacity of the HCO\(^{+}\) (1-0) line would be about two times larger than that of HCN (1-0), which is not observed in the isotope ratios (Table \ref{ap}). The derived moderate opacities of HCN (1-0) and HCO\(^{+}\) (1-0) are similar to those in the central region of the starburst galaxy NGC 253 \citep{Mei15}.

\subsection{C\(_2\)H (1-0)}\label{APc}

The intensity ratio of the triplets of the C\(_2\)H (1-0) line in LTE is expected to be \(R_\mathrm
{C_2H}\equiv\mathcal{S}_\mathrm{max}(J=3/2-1/2)/\mathcal{S}_\mathrm{max}(J=1/2-1/2)=2\) in the optically thin and \(R_\mathrm{C_2H}=1\) in the optically thick case \citep{TKT74}. From Table \ref{tab1}, the intensity ratio in the CND is measured to be \(R_\mathrm{C_2H}=2.26\pm0.19\), close to the LTE expectation, suggesting that the line is optically thin. Similar intensity ratios are also obtained toward the other regions from Table \ref{tab2}, indicating optically thin (or moderately optically thick medium): \(2.07\pm0.46\) (R1), \(1.90\pm0.41\) (R2), \(1.65\pm0.30\) (R3), \(1.57\pm0.22\) (R4), and \(1.36\pm0.31\) (R5). Applying Equation \ref{opt}, where \(\mathcal{R}=2\), and assuming that a single excitation temperature \(T_\mathrm{ex}\) holds for all hyperfine components, the optical depths of the \(J=3/2-1/2\) components in the regions R1-R5 are \(<1.0\), \(0.24\) (\(<1.4\)), \(0.89\) (within \(0.1-2.1\)), \(1.1\) (within \(0.5-2.1\)), and \(2.0\) (within \(0.8-5.9\)), respectively, where the values in brackets are uncertainties that arise from the ratio uncertainties.

\begin{table}
\begin{center}
\caption{Dense Gas Parameters in the CND Derived Under LTE}\label{ap}
\begin{tabular}{lcc}
\tableline\tableline
Parameter & HCN &  HCO\(^{+}\) \\
\hline
\(\mu\) [D] & 2.984 & 3.888 \\
\(B\) [GHz] & 44.31597 & 44.59442 \\
\(\Delta V\) [km s\(^{-1}\)] & \(137\pm4\) & \(136\pm4\) \\
\(\tau\) & 3.76 & 2.65 \\
\(N\) [cm\(^{-2}\)] (\(T_\mathrm{ex}=31\) K) & \((5.39\pm0.16)\times10^{16}\) & \((2.19\pm0.06)\times10^{16}\) \\
\(N\) [cm\(^{-2}\)] (\(T_\mathrm{ex}=10\) K) & \((6.74\pm0.20)\times10^{15}\) & \((2.75\pm0.08)\times10^{15}\) \\
\tableline
\end{tabular}
\end{center}
\end{table}

\end{document}